\documentclass[aps,pre,twocolumn,showpacs,floatfix]{revtex4}
\usepackage{graphicx}
\usepackage{subfigure}
\usepackage{epsfig}
\usepackage{amsmath}
\usepackage{graphics}
\usepackage{latexsym}
\usepackage{subfigure}

\begin{document}

\title{Equilibrium Fluid-Crystal Interfacial Free Energy of Bcc-Crystallizing Aqueous Suspensions of Polydisperse Charged Spheres}

\author{Thomas Palberg$^1$, Patrick Wette$^{2,3}$, and Dieter M. Herlach$^2$}

\affiliation{$^1$Institut f\"ur Physik, Johannes Gutenberg Universit\"at Mainz, 55099 Mainz, Germany\\
$^2$Institut f\"ur Materialphysik im Weltraum, Deutsches Zentrum f\"ur Luft- und Raumfahrt (DLR), 51147 K\"oln, Germany\\
$^3$Space Administration, Deutsches Zentrum f\"ur Luft- und Raumfahrt (DLR), 53227 Bonn, Germany}

\begin{abstract}
The interfacial free energy is a central quantity in crystallization from the meta-stable melt. In suspensions of charged colloidal spheres, nucleation and growth kinetics can be accurately measured from optical experiments. In previous work, from this data effective non-equilibrium values for the interfacial free energy between the emerging bcc-nuclei and the adjacent melt in dependence on the chemical potential difference between melt phase and crystal phase were derived using classical nucleation theory (CNT). A strictly linear increase of the interfacial free energy was observed as a function of increased meta-stability. Here, we further analyze this data for five aqueous suspensions of charged spheres and one binary mixture. We utilize a simple extrapolation scheme and interpret our findings in view of Turnbull's empirical rule. This enables us to present the first systematic experimental estimates for a reduced interfacial free energy, $\sigma_{0,bcc}$, between the bcc-crystal phase and the coexisting equilibrium fluid. Values obtained for $\sigma_{0,bcc}$  are on the order of a few $k_BT$. Their values are not correlated to any of the electrostatic interaction parameters but rather show a systematic decrease with increasing size polydispersity and a lower value for the mixture as compared to the pure components. At the same time, $\sigma_0$ also shows an approximately linear correlation to the entropy of freezing. The equilibrium interfacial free energy of strictly monodisperse charged spheres may therefore be still greater.
\end{abstract}

\maketitle

\section{Introduction}
Like their atomic counter-parts, colloidal clusters bridge between the realms of individual particles and of macroscopic bulk phases. New scale related features appear in their structural and dynamic properties and challenge their definition and description by well known purely macroscopic, colloidal or quantum concepts. Recently in this intensively studied field, much progress has been made by combining complementary approaches like analytical theory and simulations or microscopy and scattering experiments. At the same time, well controlled model systems became available. Their tuneable interactions are in many cases accurately described by analytical expressions. Because of this, we are now aware of a large variety of different colloidal cluster types with different shape, internal structure, dynamics and cluster-cluster interaction. Their formation, stability and properties are of fundamental interest \cite{Wales Book 1999, Baletto RMP 2005,Glotzer Nat Mat 2007, Malins JPCM 2009, Li Angew Chem 2011, Calvo Nanoscale 2012, Kraft PRE 2013} but moreover play a decisive role in practical processes, like gelation, coating or food processing \cite{Lu Annu rev 2013, Weis J Rheo 2014}. Clusters also appear as embedded ensembles in colloidal (shear-)melts during the initial steps of freezing \cite{HJS PRL 2006, Herlach JPCM 2011 colloids as models} and vitrification \cite{HJS JNCM 2002}. Unlike in condensation problems, here the differences between cluster and melt properties (e.g. density or structure) may become very small and require special care in cluster discrimination \cite{ten Wolde Science 1997, Gasser JPCM 2003, Lutsko PRL 2006}. Moreover, cluster and interfacial structure may be time and/or size dependent \cite{Tan Nat Phys 2014, Kratzer SM 2015, ten Wolde PRL 1995, Moroni PRL 2005}. Cluster energetics, dynamics and growth kinetics determine the polymorph selection and the emerging solid's micro-structure \cite{Russo Sci Rep 2012, Tan Nat Phys 2014, Russo SM 2012} but furthermore may make them a crystallization frustrating agent in vitrifying melts \cite{HJS JNCM 2002, Tarjus JPCM 2005, Royall Nat Mat 2008, Franke SM 2014}. In this instance, their local orientation prevents coalescence and their local symmetry hinders a global transformation to the crystalline state \cite{Bernal Nature 1962, Anikeenko PRL 2007}. Clusters are therefore considered to be an important transient state in the formation of (colloidal) solids.
Both simulations and microscopy reveal the shape of such nuclei to be anisometric and their surface to be rough, extended and (as expected by scale arguments) not well definable in terms of continuous, differentiable two dimensional manifolds \cite{Schilling PRL 2011, kawasaki PNAS 2010, Russo Sci Rep 2012, Tan Nat Phys 2014, Kratzer SM 2015}. Light scattering \cite{harland PRE 55, Wette EPL 2003, Wette JCP 2005 binary, HJS PRL 2006, Kozina SM 2014, Franke SM 2014} and small angle x-ray scattering experiments \cite{Herlach Rev EPJST 2014} return the statistically well founded temporal development of the orientationally averaged cluster sizes, the cluster numbers and their rate of production \cite{harland PRE 55, Palberg JPCM Rev 1999, Gasser JPCM 2009, Palberg JCPM Rev 2014}. Given their accessibility by complementary methods working on different length scales, colloidal clusters appear to be well-suited models for detailed studies of phase transformation processes as well as critical assessments of the concepts employed in their description.

The key quantity of interest in the present paper is the reversible work involved in the creation of a dividing surface, an interface between a cluster and its surroundings. It is termed surface tension, interfacial tension or interfacial free energy (IFE). On the macroscopic level with adjacent continuous phases, this is a well-defined quantity and it can be determined with high accuracy by theoretical, numerical and experimental techniques. On the level of clusters, a statistically meaningful description can only be based on orientationally averaged quantities. To connect this data to the desired (equilibrium) thermodynamic quantities, no generally accepted scale bridging theory is available. Thus very often approximate and empirical schemes are used to parameterize the raw data. A widely used parametrization for crystallization processes is the so-called classical nucleation theory (CNT)\cite{Volmer ZPhys 1926, Kaischew ZPC 1934, Becker An Phys 1935, Zeldovic JETP 1942, Turnbull JCP 1949, Lutsko JCP 2012}. Since it basically ignores the cluster nature of nuclei and takes a macroscopic view of this microscopic problem, conceptual difficulties have been pointed out. (The interested reader is referred to Appendix A, where she finds a short outline of the main objections raised and some suggestions to circumvent these). Still this approach has turned out to be eminently practical and versatile. It has also opened a way to obtain estimates of CNT-based effective IFEs from non-equilibrium crystallization experiments on systems as diverse as metals and colloids, hard spheres (HS) and Lennard Jones particles. CNT therefore became central in modeling the kinetics of first order phase transitions \cite{Jiang rev Surf Sci Rep 2008, Antonovicz Rev Adv Mat Sci 2008, MRS 29 2004, Kelton Solid state Phys 1993}. In polymorph-selection for instance, CNT-based kinetic arguments suggest, that the cluster with the lowest nucleation barrier will reach its critical size and continue to grow \cite{Ostwald Z phys Chem 1897, Alexander McTague PRL 1978}, rather than that of the lowest free energy. CNT also found important practical application, e.g. in the fabrication of advanced soft materials \cite{Dziomkina SM 2005, Gorishnyy PRL 2005, Sear JPCM Rev Nucl 2007}. Moreover, intriguingly simple empirical rules have been discovered applying CNT to crystallization phenomena. This paper takes particular interest in Turnbull's rule relating the IFE per particle in a surface to the latent heat of fusion per particle \cite{Turnbull JAP 1950} or the entropy of fusion per particle \cite{Laird JCP 2001} in a linear fashion.

In fact, it is with this rule in mind, that in the present paper we undertake a comprehensive analysis of presently available CNT-based estimates of effective non-equilibrium IFEs for several systems of polydisperse charged sphere suspensions. The original data was obtained by optical experiments \cite{Wette EPL 2003, Wette JCP 2005 binary, Wette JCP 2005 Microscopy,Wette PRE 2007 CNT, HJS Engelbrecht SM 2011, Herlach JPCM 2011 colloids as models} yielding nucleation rate densities via KJMA-theory \cite{Avrami, Johnson_Mehl, vanSiclen} or Kashchiev's theory of transient nucleation \cite{kashchiev}. Nucleation rate densities in turn were parameterized using CNT \cite{Wette PRE 2007 CNT, Herlach JPCM 2011 colloids as models} with the independently measured melt meta-stability expressed in terms of the chemical potential difference between the two phases, $\Delta \mu$, as input \cite{Aastuen PRL 1986, Wurth PRE 1995}. This yielded kinetic pre-factors and the CNT-based effective non-equilibrium IFEs, $\gamma$, used as the starting point of the present analysis.

In our analysis, we go beyond previous work, as we apply a simple extrapolation scheme to obtain first estimates of  CNT-based effective equilibrium IFEs for bcc crystallizing model systems. We compare these to values for equilibrium IFEs obtained for various systems by direct observation of equilibrated macroscopic interfaces. We also compare to non-equilibrium IFEs, both CNT-based effective IFEs for atomic and molecular systems and more directly obtained ones from e.g. direct observations of cluster fluctuations. Our comparison to this data from both experiments and simulations reveals that the IFEs of polydisperse charged colloidal systems range between those of metals and of monodisperse hard spheres but are much larger than those found for point Yukawa systems. Next we search for correlations of the inferred CNT-based effective equilibrium IFEs with the system specific parameters characterizing strength and range of the prevailing electrostatic interactions as well as to other properties of the cluster constituents, e.g. their colloid specific polydispersity. Interestingly we observe no correlations to the former quantities. However, we do observe a pronounced anti-correlation of the IFE to the system polydispersity. We further propose an extension of the extrapolation scheme based on Turnbull's relation, some elementary thermodynamics and the assumption that the entropy of freezing doesn't depend on the degree of meta-stability. This procedure returns estimates of other thermodynamic quantities like the entropy of freezing, the latent heat of freezing and Turnbull's coefficient. Our findings allow rationalization of the observed anti-correlation of the IFEs to the system polydispersity in terms of the entropy differences between the adjacent structures. They further support entropy based theoretical arguments for the dependence of Turnbull's coefficient on crystal structure.

Data for comparison comes from different experimental and theoretical approaches on a large variety of systems. Absolute values of the IFE differ by orders of magnitude due to the different particle number densities, $n$, involved for e.g. metals (${n\approx10^{26} m^{-3}}$) and colloids (${n\approx (10^{17} - 10^{19}) m^{-3}}$). In view of this, we follow the original work of Turnbull \cite{Turnbull JAP 1950} and normalize $\gamma$  with the area taken by a single particle in the interface, $A_P$, to compare reduced values of the interfacial free energy, ${\sigma=\gamma A_P}$. In the literature, different measures for $A_P$ have been employed. For metals \cite{Turnbull JAP 1950} and hard spheres (HS) \cite{harland PRE 55, Auer Nature 2001 HS mono, Hartel PRL 2012 HS}, but sometimes also for strongly screened charged spheres \cite{Auer JPCM 2002 weakly charged}, $A_P$ was approximated by $(2a)^2$, where $a$ is the particle radius. This is generally considered as a physically reasonable approximation, since at the large volume fractions encountered in close packed metals, HS and slightly charged HS crystals the particles are (nearly) in contact. Further, any change in density with increased meta-stability is generally small, such that $A_P\cong const.$ for all $T$ below the melting temperature $T_M$, respectively all volume fractions above the freezing volume fraction ($\Phi_{F,HS}=0.495$ \cite{zykova-Timan JCP 2010 HS AHS}). By contrast, in low salt charged sphere (CS) crystals the nearest neighbour distance at melting is usually on the order of several particle diameters due to mutual electrostatic repulsion ($\Phi_F\leq0.01$ \cite{Monovoukas JCIS 1989}; see also Tab I, below). Here, the area of interest is the square of the nearest neighbour distance, ${d_{NN}} ^2$. Note that this area will shrink considerably when the particle number density is increased above the melting density. For the here analyzed samples the spread in ${d_{NN}}^2$ covers about three orders of magnitude between its value at melting for PnBAPS70 ($n_{M,PnBAPS70} = 2\mu m^{-3}$) and the one at the largest investigated particle concentration for Si77 ($n_{max,Si77} = 80\mu m^{-3}$). We therefore normalize each non-equilibrium IFE by the square of the nearest neighbour distances at the particle number density investigated  ${d_{NN}} ^2 = n^{-2/3}$. We further use $k_BT_M$ as energy unit for the reduced IFEs, where $T_M$ denotes the melting temperature.

Important reference data for our comparison comes from studies of equilibrated interfaces. Here, previous studies focused on systems where the melt is in contact with a close-packed crystal structure. Equilibrium IFEs, stiffness and anisotropy have been studied for theoretical model systems like hard spheres (HS) \cite{Hartel PRL 2012 HS, zykova-Timan JCP 2010 HS AHS, davidchack JPCB 2005 MD HS, Mu JPCB 2005 anisotropy HS, davidchack JCP 2006 fluct HS, Amini PRB 2008 fluct HS, davidchack JCP 2010 HS,  Fenandez PRL 2012 HS}, Lennard Jones (LJ) particles or particles interacting via inverse power potentials \cite{davidchack JCP 2003 LJ, davidchack PRL 2005 1/r, davidchack JPCB 2005 1/r, benjamin JPC 2014 LJ WCA, Turci EPJST 2014}. These quantities have also been studied in simulations of various metal systems utilizing embedded atom potentials \cite{Johnson JMR 1989, Foiles PRB 1990}. Experimental equilibrium studies are rare for metal systems due to the practical difficulties involved in working at the melting point \cite{MRS 29 2004, Egry EPJST 2014}. Equilibrated interfaces have, however, been studied experimentally in HS colloidal systems \cite{HS}. Hern\'{a}ndez-Guzm\'{a}n and Weeks, for instance, performed a capillary wave analysis of the equilibrated interface between a face centered cubic (fcc) crystal of HS and the adjacent HS fluid \cite{Hernández-Guzmán PNAS 2009}. Rogers and Ackerson measured the IFE for HS crystals from a careful groove analysis of a HS polycrystal-fluid interface \cite{Rogers Phil Mag 2011}. They obtained a value of $\sigma_{0,HS} = (0.58 \pm 0.05) k_BT$ for the reduced equilibrium IFE. This value is in close agreement with theoretical expectations and simulation results which - depending on the approach taken - give orientationally averaged values of the reduced equilibrium IFE, $\sigma_{0,HS} = (0.56-0,68)k_BT$ \cite{davidchack JPCB 2005 MD HS, Mu JPCB 2005 anisotropy HS, davidchack JCP 2006 fluct HS, Amini PRB 2008 fluct HS, davidchack JCP 2010 HS, zykova-Timan JCP 2010 HS AHS, Hartel PRL 2012 HS, Fenandez PRL 2012 HS}.

More recent work also addressed body centred cubic (bcc) crystal structures in contact with their melt. Heinonen et al. studied crystallizing point Yukawa systems, comparing state-of-the-art molecular dynamics simulations and theoretical approaches \cite{Heinonen JCP 2014}. For this particular kind of long ranged repulsive electrostatic interaction, the authors obtained IFEs which were much lower than those of HS. Also for bcc crystallizing metals, a few simulation studies have been reported \cite{Hoyt MRS 2004}. In these studies, IFE values are on the same order as those of fcc or hexagonal close packed (hcp) metals \cite{Hoyt MRS 2004, Rozas EPL 2011}, but their IFE anisotropy and temperature dependence as well as Turnbull's coefficient are predicted to be considerably smaller than for fcc metals.

A second set of data was obtained in studies on non-equilibrium clusters using CNT to extract effective IFE values from their nucleation rate densities. Data is available for both atomic systems \cite{Turnbull JCP 1949, Turnbull JAP 1950, MRS 29 2004, Kelton Solid state Phys 1993, Sear JPCM Rev Nucl 2007, Klein PRB 2008 Zr} and colloids \cite{harland PRE 55, Palberg JPCM Rev 1999, Gasser Science 2001, Auer Nature 2001 HS mono, Auer Nature 2001 polydisperse, Auer JPCM 2002 weakly charged, Auer AnnRevPC 2004, Gasser JPCM 2009, Palberg JCPM Rev 2014, Franke SM 2014}. The non-equilibrium CNT-based effective IFEs for the hard sphere colloids agree with the values obtained under equilibrium conditions with reduced values of about ${\sigma\cong 0.55k_B T}$. This is particularly true for those values obtained at the melting volume fraction \cite{Franke SM 2014}. Concerning metals, we note that experimental data obtained at the nucleation temperature of metals, $T_N$, may be converted to estimates of the effective IFEs at the equilibrium melting temperature, $T_M$. This has been demonstrated e.g. in the case of Ni in a combined study of calorimetrically obtained nucleation rates and state-of-the-art simulations \cite{bokeloh PRL 2015}. There, a near quantitative agreement of the estimated values with the values measured for the same system at equilibrium was observed \cite{Rozas EPL 2011}. To perform these estimates and conversions for systems with dominant hard-core interaction, a constant entropy of freezing appears to be a sufficient assumption \cite{Laird JCP 2001}. More sophisticated corrections including size dependence of the IFE and the temperature dependence of the enthalpy of freezing have been discussed in \cite{Jiang rev Surf Sci Rep 2008}. This large data compilation further suggests that Turnbull's coefficient remains unaffected by the mentioned conversion, i.e. it can equally well be read from the reduced non-equilibrium IFE at the nucleation temperature and the equilibrium IFE at the melting temperature. Interestingly, the CNT-based effective non-equilibrium IFEs, the interfacial stiffnesses, and their temperature dependencies are found to be much smaller in the bcc than in the fcc case, when compared for the same interaction type and strength. Also, in a recent extensive simulation of nucleation in hard-core Yukawa systems the non-equilibrium IFE was found to be a factor two smaller in the case of bcc crystals as compared to the case of fcc crystals \cite{Kratzer SM 2015}. This observation is also theoretically supported within the negentropic model of Spaepen et al. \cite{Spaepen Acta Metallica 1975, Spaepen Scr Metallica 1976} and the broken bond model of Gr\'{a}n\'{a}sy and Tegze \cite{Granazy MDF 1991 broken bond model}. Thus, also Turnbull's coefficient $C_T$ should take different values for different structures. In fact, experimental values for $C_T$ of fcc crystallizing metals converge to $C_{T,fcc} = 0.43$ \cite{Kelton Solid state Phys 1993} while simulation results for fcc crystallizing metals are better described by $C_{T,fcc} = 0.55$ \cite{Hoyt MRS 2004}. The few simulation results available for bcc crystallizing metals are best described by $C_{T,bcc} = 0.29$ \cite{Hoyt MRS 2004}. Our first systematic estimate of ${C_{T,bcc,exp} = 0.31\pm0.03}$ appears to be much smaller than the values found for fcc crystallizing systems and in is remarkably close agreement with predictions for bcc systems.

The comparison of theoretical and experimental results faces yet another difficulty in that the bulk of theoretical IFE studies focused on strictly monodisperse systems. By contrast, all experimental studies on colloids have to cope with an inevitable polydispersity, characterized by a polydispersity index, $PI = s_a/\bar{a}$, where $\bar{a}$  is the mean particle radius and $s_a$ is the standard deviation. Data considered in the present paper was taken on samples of different PIs ranging between 0.025 and 0.08. Interestingly, in their simulations on the crystallization of polydisperse HS, Auer and Frenkel found both the (expected) decrease of the nucleation barrier with increased meta-stability but moreover observed an increase with increased polydispersity. This increase occurred for $PI > 0.05$ and was interpreted in terms of an increase of IFE with increasing PI \cite{Auer Nature 2001 polydisperse}. On the experimental side, Sch\"ope et al. reported a dramatic slowing of the onset of nucleation, but at the same time also an increase in nucleation rate densities for HS systems, when the PI was increased by a mere percent from 0.048 to 0.058 \cite{Schope PRE 2006 polyd}. These authors discussed their findings as indicative of the onset of fractionation processes. The latter are expected to occur for PI $\geq$ 0.05 \cite{Fasolo PRL 2003, Fasolo PRE 2004, Sollich PRL 2010, Sollich SM 2011} and are also observed for eutectic binary mixtures of HS and attractive HS \cite{Kozina SM 2014, Ganagalla JCP 2013, Kozina SM 2012, Fasolo JCP 2005}. They are also predicted and observed for charged particles but at much larger PI \cite{VanderLinden JCP 2013, Labbez 2015 preprint}. Therefore, the PI seems to have a strong influence on nucleation kinetics. But the important open question remains, how IFEs react to polydispersity particularly at comparably low values of PI, where fractionation effects are expected to play a subordinate role. Having access to a large number of systems differing with respect to polydispersity, we are able, for the first time, to look at its influence in a more systematic way. We find that the IFE systematically decreases with increasing polydispersity, while the Turnbull coefficient remains unaffected.

The remainder of the paper first will quickly recall the characteristics of the investigated samples, the experimental procedures and the evaluation schemes leading to the reported non equilibrium IFEs. We then will determine the $\sigma_{0,bcc}$ for the five pure species and one binary mixture from a simple extrapolation scheme. In an extension of this scheme, we will further estimate other thermodynamic quantities including Turnbull's coefficient. We continue with an extensive discussion of our findings, where we address the observed values and their spread, the observed anti-correlation of $\sigma_0$ to the PI and the observed Turnbull coefficients for bcc crystallizing systems. After that we will give our conclusions. There are several appendices that provide more background information on: A) additional correlation checks; B) the characterization of particle interactions under deionized conditions; C) the determination of CNT-based effective non-equilibrium IFEs from nucleation and growth measurements; and D) the use of CNT and related schemes to obtain non-equilibrium IFEs.

\section{Analyzed systems and their characteristics}
\subsection{Particle characterization}
We start with shortly recalling the characteristics of the investigated systems and the methods employed in obtaining the Non-equilibrium IFE. A more detailed discussion of the experiments and raw data interpretation can be found in Appendix C. We analyze data of five species of moderately to highly charged colloidal spheres in aqueous suspension and one binary mixture. Co-polymer particles were kindly provided by BASF, Ludwigshafen. Silica-particles were home made employing St\"ober-synthesis. Systems under consideration are compiled in Tab.~I with the corresponding references for the measurements of the nucleation rate densities \cite{Wette EPL 2003, Wette JCP 2005 binary, Wette JCP 2005 Microscopy, Wette PRE 2007 CNT, HJS Engelbrecht SM 2011, Herlach JPCM 2011 colloids as models}. Sample lab codes refer to the particle material (Polystyrene: PS; Poly-N-Butylacrylamide: PNBA; Silica: Si) and particle diameter (in nm).

For the present investigation, size characterization is of prime importance. Si and PS diameters were obtained from Transmission Electron Microscope (TEM) images with the PS data quoted from the manufacturer. Ultra small angle x-ray scattering (USAXS) form factor measurements on the Si particles gave coincident values. PS particles were further investigated by static light scattering returning slightly larger values for the geometric radii, indicating a slight shrinkage of particles under TEM-conditions. They were also investigated by dynamic light scattering returning hydrodynamic radii which are larger than the geometric radii by some 5\%. The co-polymer particles are not stable under the TEM. Their diameters were determined by the manufacturer from ultracentrifugation. Here, static light scattering gave slightly smaller diameters (by about 1.5\%), dynamic light scattering again gave some 5\% larger values. While this spread in diameters reflects the (known) differences between the applied methods \cite{Lange PPSC 1995}, this study is primarily interested in the corresponding polydispersities. Using TEM, some 1500 particles were counted for each species. Form factors from USAXS and static light scattering were analyzed using a polydisperse Mie-fit \cite{ballauff CurrOpn 2001}, dynamic light scattering was analyzed using the cumulant method. In each case the statistical uncertainties of the standard deviation of the diameters are estimated to be below 10\%, with the lowest uncertainties for the analytical ultracentrifugation and the largest for the cumulant method. However, for ultracentrifugation an additional systematic uncertainty arises from the use of the bulk PnBAPS-co-polymer density, as does for static light scattering from modelling the particles as homogeneous spheres utilizing the bulk index of refraction. The combined statistic and systematic uncertainty in the PI are therefore dependent on the choice of method to determine the average diameter. A conservative estimate gives 15\% uncertainty as an upper limit for the PIs shown in Tab. I. Interestingly, however, for each species, the PI-values obtained by different methods agreed within 5 to 7 percent.
\begin{table*}[ht]
\begin{tabular}{|c|c|c|c|c|c|c|c|c|c|} \hline
Sample &  Refe- & 2a$/nm$ & PI & $Z_{eff,G}$ & $\Psi_{eff}$ & $n_F/\mu m^{-3}$ & $n_M/\mu m^{-3}$& $d_{NN}/\mu m$& $d_{NN}/(2a)$  \\
Batch No. & rence & & & & & & & at melting & at melting \\ \hline
PNBAPS68 & \cite{Wette EPL 2003, Wette JCP 2005 Microscopy, Wette PRE 2007 CNT}  & 68 & 0.05 & 331$\pm$3 & 9.5 & 6.0$\pm$0.3 & 6.1$\pm$0.3& 0.55&8.1\\
BASF ZK2168/7387 & & (UZ) & & & & & & & \\ \hline
PNBAPS70  & \cite{HJS Engelbrecht SM 2011} & 70 & 0.043  & $325\pm 3$ & 8.6 & 1.8$\pm$0.2 & 2.0$\pm$0.2 &0.79 &11.2\\ 
BASF GK0748 & & (UZ) & & & & & & & \\ \hline
SI77 & \cite{Herlach JPCM 2011 colloids as models} & 77 & 0,08 & 260$\pm$5 & 6.4 & $>$ 28$\pm$1 & 30$\pm$1&0.32 & 4.1\\
&&(TEM) & & & & & & &\\ \hline
PS90  & \cite{Wette JCP 2005 binary} & 90 & 0.025 & 315$\pm$8 & 8.1 & 4.0$\pm$0.5 & 4.6$\pm$0.5&0.60 &6.7\\ 
Bangs Lab 3012 & & (TEM) & & & & & & &\\ \hline
PS100B  & \cite{Wette JCP 2005 binary}& 100 & 0.027 & 327$\pm$10 & 7.6 & 4.2$\pm$0.5 & 5.5$\pm$0.2 &0.57&5.7\\ 
Bangs Lab 3067 & &(TEM) & & & & & & &\\ \hline
\end{tabular}
\medskip
\caption{Suspension data: Lab code and/or manufacturer's Batch No.; references for the kinetic data; diameter with experimental method indicated: UZ: Ultracentrifuge measurements performed by the manufacturer, TEM: Transmission Electron Microscopy; polydispersity index PI (standard deviation normalized by mean diameter, values given refer to diameters and standard deviations measured with the methods indicated. Additional measurements employing USAXS form factor measurements, static light scattering and dynamic light scattering gave values agreeing within some (5-7)\% with the values quoted; see text for further details); effective charge number, $Z_{eff,G}$ from elasticity measurements; dimensionless effective surface potential, $\Psi_{eff}$, assuming the effective charge saturation limit; freezing (F) and melting (M) number densities from static light scattering; nearest neighbour spacing $d_{NN}$ at melting in microns; and in terms of the particle diameter.}
\end{table*}

Size-polydispersities range from low (PI = 0.025) to moderate (PI = 0.08). These values are the geometric polydispersities, which hardly alter the inter-particle spacing and thus the strength of pair interactions. For charged spheres, however, one expects an additional charge-polydispersity, altering the pair interaction at $d_{NN}$. For highly charged low salt systems, like the ones investigated, a suitable analytic expression for the effective pair interaction is given by a Debye-Huckel pair potential with a renormalized effective charge \cite{Alexander JCP 1984, Gisler JCP 1994, Levin RPP 2002, Shapran CSA 2005} which scales linearly with the particle radius \cite{Garbow JPCM 2004} (see also Appendix B). Thus any size-polydispersity directly translates into a charge-polydispersity. However, little is known about the effects of charge-polydispersity on the properties of charged sphere suspensions, with the exception that effects on the phase behaviour are smaller in CS than in HS systems with the same PI \cite{Sollich JPCM 2002}. In fact, a procedure proposed by L\"owen et al. \cite{Lowen JPCM 1991} to map charge-PIs to HS size-PIs gave some 50\% smaller values in the latter case. Unfortunately, this procedure is not reliably applicable for the present highly charged low salt systems \cite{Lowen priv Comm}. In a recent study, van der Linden et al. showed that the maximum PI compatible with crystal formation without fractionation is about 0.13 \cite{VanderLinden JCP 2013} which is to be compared to the maximum PI of about  0.062 in the HS case \cite{Fasolo PRE 2004}. Similarly, no fractionation effects are observed for any of our single component systems and furthermore the mixture crystallizes as bcc substitutional alloy with a spindle type phase diagram \cite{Lorenz JCP 2010}, whereas binary mixtures with HS-like interactions would form an eutectic at this geometric size ratio \cite{Geerts JPCM 2010}. We therefore note, that the given PIs should be taken as upper limit and that the effective PIs are probably much smaller.

All particles are negatively charged and were investigated under thoroughly deionized conditions using advanced, continuous ion exchange techniques \cite{Wette JCP 2001 Deionization}. For further details of the preparation procedure and the characterization of the interaction strength and range under deionized and strongly interacting conditions, the interested reader is referred to Appendix B. We assume to have the highly charged particles in or close to the effective charge saturation limit \cite{Levin RPP 2002, Shapran CSA 2005} and calculate the dimensionless effective surface potential, $\Psi_{eff} = Z_{eff,G} \lambda_B / a$ as a measure of interaction strength \cite{Alexander JCP 1984, Gisler JCP 1994}. Here $a$ is the particle radius, $\lambda_B = 0.72nm$ is the Bjerrum length in water and $Z_{eff,G}$ is the effective charge from elasticity measurements \cite{Wette CSA 2003}. In the deionized state, all samples including the mixture form polycrystalline bcc solids for $n \geq n_F = (2-8)\mu m^{-3}$ with the location of the melting line coinciding with theoretical expectations \cite{Lorenz JCP 2010, Wette JCP 2010 PDG Si, Wette PCPS 2006 Consistence}.

\subsection{Crystallization experiments and determination of non-equilibrium IFEs}
Use of the continuous deionization technique cycling the suspension in a closed tubing circuit keeps the systems in a homogenized shear-molten state prior to re-solidification \cite{Wette JCP 2001 Deionization}. Growth measurements were performed in rectangular cells with the growing wall crystals observed by Bragg microscopy \cite{Palberg JPCM Rev 1999}. For all samples and $n > n_M$, growth was observed to be strictly linear in time, characteristic of reaction limited growth. The number density dependent growth velocity, $v$, was interpreted in terms of a Wilson Frenkel growth law to obtain an estimate of the difference in chemical potential between melt and crystal, $\Delta \mu$. This was done for PnBAPS68, PnBS70, PS100B and Si77 by fitting a modified Wilson-Frenkel expression which, following W\"urth, was based on a reduced energy density difference calculated with a density dependent interaction potential \cite{Wurth PRE 1995}. For PS90, no growth data was available at the time of publication of the nucleation rate densities \cite{Wette JCP 2005 binary}. Therefore a simple estimate for $\Delta \mu$ was used based on the reduced density difference: ${\Delta \mu = B (n - n_F)/n_F}$ with the proportionality constant reported by Aastuen et al. for particles of 91nm size: $B = 10$ \cite{Aastuen PRL 1986}. Later, measurements of the growth velocity in dependence on particle concentration were made for PS90. The modified Wilson-Frenkel-fit of this data using Aastuen's expression returned a proportionality constant of ${B_{PS90} = (4\pm0.6)}$ (\cite{Lorenz unpublished}, see also Appendix C). Since also the evaluation of PS100B growth data using Aastuen's approximation yielded ${(B=4.0\pm0.2)}$, we choose to adapt a value of ${B=4}$ for the mixture, too.

\begin{figure}[htb]
\begin{center}
\includegraphics[width=0.50\textwidth,angle=0]{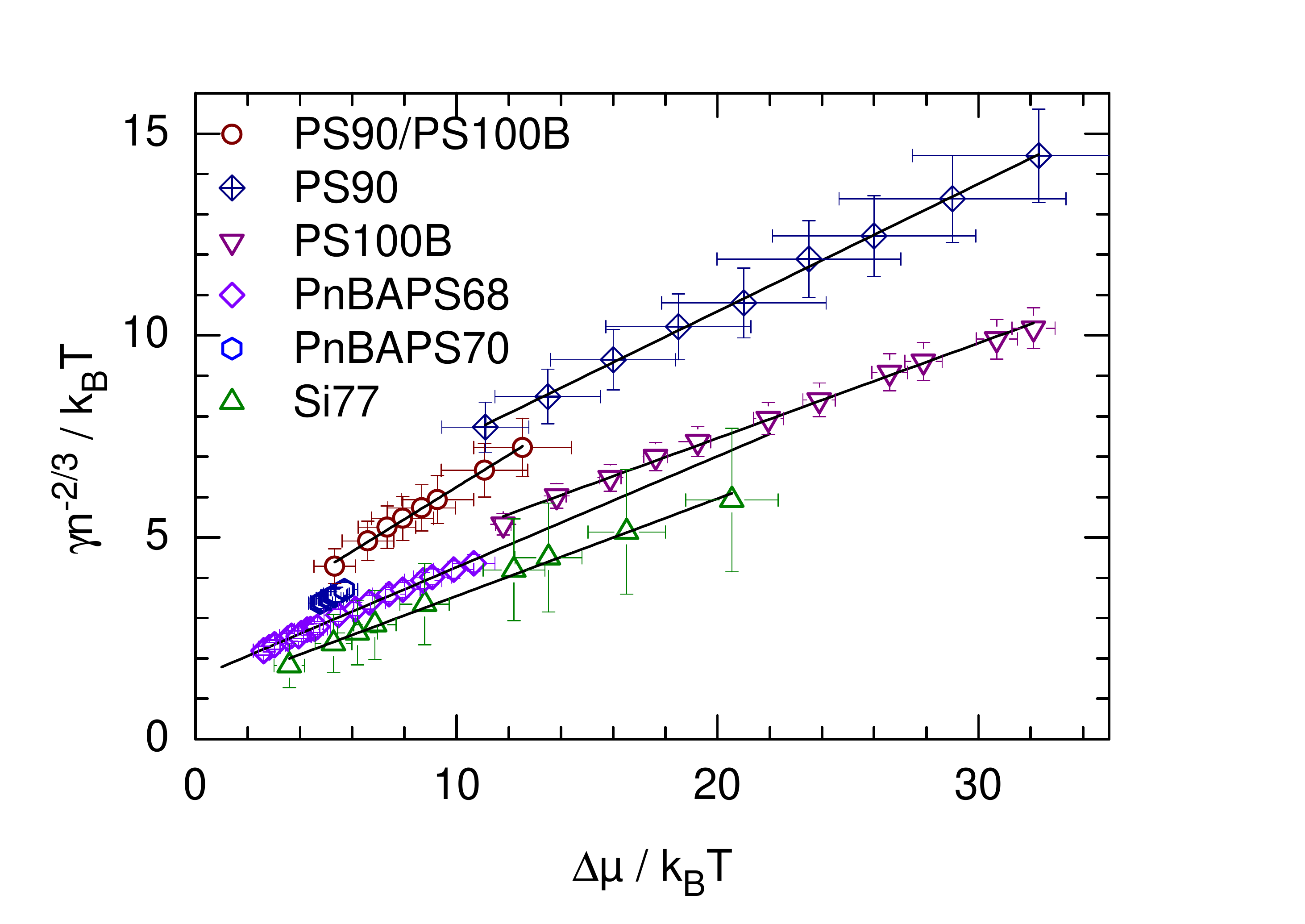}
\caption{\label{fig1}(Color online) Dependence of the reduced interfacial free energies on meta-stability, $\Delta \mu$, for the indicated species. Interfacial free energies $\gamma$  (in $J/m^2$) were normalized by $n^{-2/3}$ and plotted in units of $k_B T$ as quoted from the original literature or re-evaluated from the original data using a corrected value for $\Delta\mu$. Solid lines are least-square-fits of $\sigma = \sigma_0 + m \Delta \mu$. Note that here and throughout, we adopt the convention to consider the crystal as educt and the melt as product, as to obtain positive values for $\Delta \mu$, $\Delta H_f$ and $\Delta S_f$.}
\end{center}
\end{figure}

The nucleation rate density, $J$, varies drastically with increasing $n$. Therefore, different techniques were employed in its study. At low densities, $J$ was determined directly from video-microscopy in Bragg-microscopic mode \cite{Palberg JPCM Rev 1999, Wette JCP 2005 Microscopy}, at medium densities \textit{via} past-solidification size counts from polarization microscopy \cite{HJS Engelbrecht SM 2011} and at high densities from Bragg scattering or time resolved USAXS measurements \cite{Wette EPL 2003, Wette JCP 2005 binary, Wette PRE 2007 CNT, Herlach JPCM 2011 colloids as models}. All three approaches yield consistent $J(n)$ for a given sample. For CS, $J(n)$ first increases near exponentially with increasing $n$, then the increase gradually slows, but no decrease like the one known from HS is observed (for a comparison see e.g. \cite{Wette JCP 2005 binary} and Appendix C). From the combined data sets of $J(n)$ and $\Delta \mu (n)$ the nucleus-melt interfacial free energies, $\gamma (n)$, were derived in the frame work of CNT. Either we performed a least-square fit of Eqn.(C.8) to the data using a proportionality constant $A$, an effective long time self-diffusion coefficient ${D_{L}^S(n)}$ and $\gamma (n)$ as free parameters, or a graphical evaluation method was utilized in which $\gamma (n)$ was determined from the local slope more directly without any assumption about the kinetic pre-factor (for details see \cite{Wette EPL 2003, Wette PRE 2007 CNT} and Appendix C). For PS90 and the mixture the original nucleation data was re-evaluated using ${B=4k_BT}$ to return slightly altered IFEs and a significantly increased dependence on ${\Delta\mu}$.

In Fig.~1 we re-plot the data reported for PnBAPS68, PnBAPS70, Si77 and PS100B together with the corrected data for PS90 and the 1:1 mixture of PS90 and PS100B in terms of their reduced values ${\sigma = \gamma(\Delta \mu){d_{NN}}^2}$. The data sets show different uncertainties. In particular for Si77, graphical evaluation lead to an enlarged statistical uncertainty in $\sigma$. Furthermore, estimating $\Delta \mu$ following Aastuen and neglecting the density dependence of the interaction potential results in an enhanced systematic uncertainty in $\Delta \mu$ for PS90 and the mixture. Note that this will influence only the slope but not the intercept of any linear fit since the freezing point with $\Delta \mu = 0$ is accurately known. For the other samples, $\gamma$ was determined from fits of classical nucleation theory expressions \cite{Wette PRE 2007 CNT} and $\Delta \mu$ was determined from growth experiments following \cite{Wurth PRE 1995}. There, the uncertainties are mainly statistical and remain on the order of a few percent.

\section{Data analysis and results}
\subsection{Equilibrium IFEs and their dependencies on system parameters}
Fig.~1 reveals that, within experimental error, the reduced non-equilibrium IFEs of all charged sphere samples show a strictly linear increase for increasing meta-stability. This suggests the use of the scheme sketched in Fig.~2 and simply extrapolate the data to zero $\Delta\mu$ without making any further assumptions. We obtain the extrapolated equilibrium IFE and the slope by performing least-square fits of $\sigma = \sigma_0 + m \Delta \mu$. The results for $\sigma_ 0$ and $m$ are displayed in Tab.~II.

\bigskip
\begin{table*}[ht]
\begin{tabular}{|l|l|l|l|l|l|l|}\hline
& PnBAPS68 & PNBAPS70 & Si77 & PS90 & PS100B & PS90/PS100  \\ \hline
$\sigma_0 / k_BT$ & 1.51$\pm$0.04 & 1.62$\pm$0.07 & 1.13$\pm$0.16 & 4.28$\pm$0.43 & 2.75$\pm$0.11 & 2.26$\pm$0.16 \\ \hline
$m = C_{T,bcc}$ & 0.274$\pm$0.003 & 0.364$\pm$0.018 & 0.254$\pm$0.008 & 0.316$\pm$0.005 & 0.235$\pm$0.005 & 0.405$\pm$0.011 \\ \hline
$\Delta S_f/J \; mol^{-1}K^{-1}$ & 1.85$\pm$0.1 & 1.51$\pm$0.13 & 1.58$\pm$0.17 & 4.56$\pm$0.47 & 3.93$\pm$0.27 & 0.57$\pm$0.05 \\ \hline
$\Delta H_f/kJ \; mol^{-1}$ & $0.55 \pm 0.3$ & $0.45 \pm 0.4$ & $1.36 \pm 0.14$ & $1.17 \pm 0.8$ & $1.17 \pm 0.8$ & $0.57\pm0.05$ \\ \hline
\end{tabular}
\medskip
\caption{Results for a fit of $\sigma = \sigma_0 + m \Delta \mu$ to the data shown in Fig.~1. We obtain the intercept, i.e. equilibrium reduced IFE, $\sigma_ 0$, and the slope $m$, i.e. the Turnbull coefficient $C_{T,bcc}$. Further, we show estimates of the the molar entropy of fusion, $\Delta S_f$, and the molar enthalpy of fusion, $\Delta H_f$.}
\end{table*}

\begin{figure}[htb]
\begin{center}
\includegraphics[width=0.35\textwidth,angle=0]{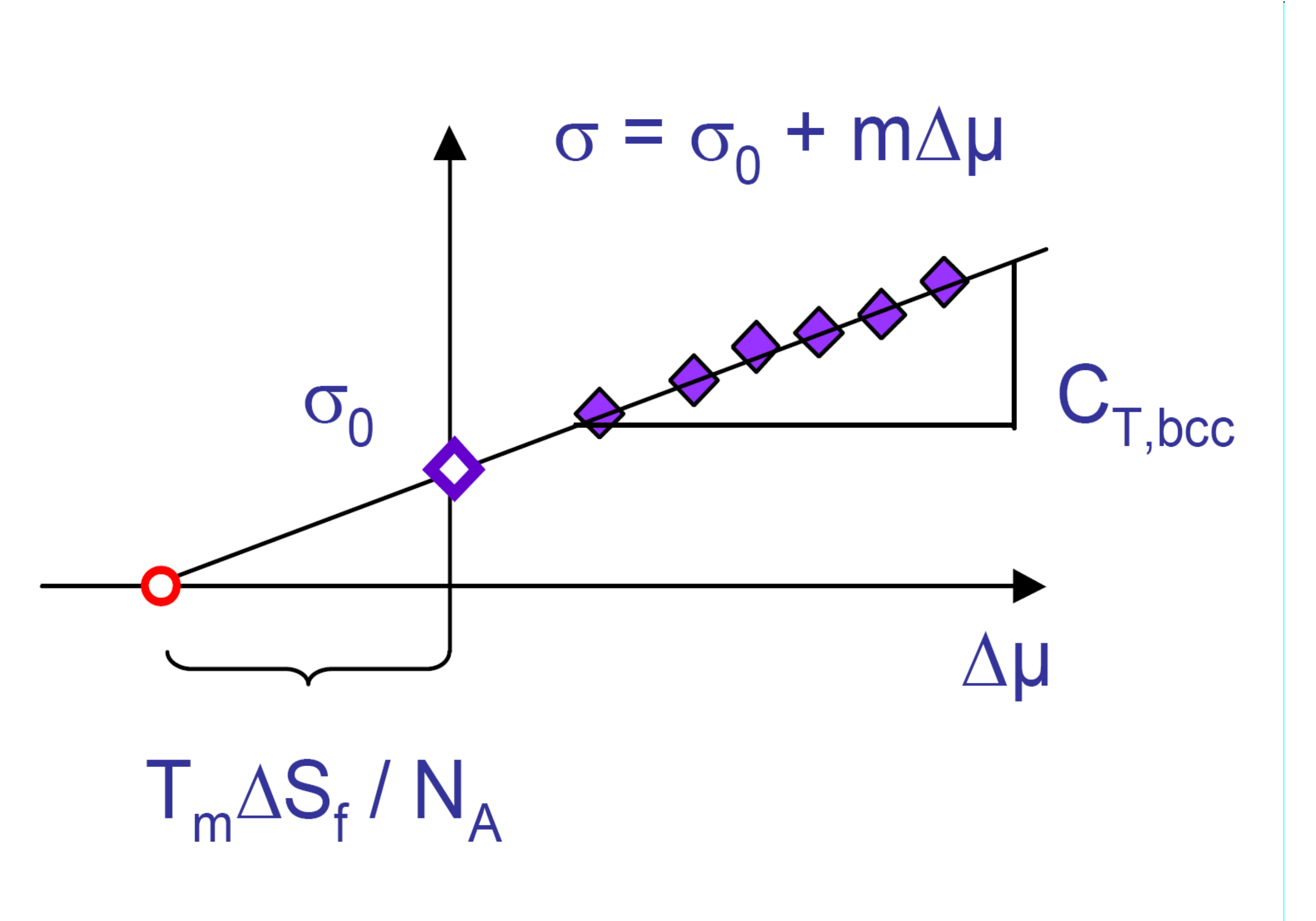}\hspace{0.1cm}
\caption{\label{fig2}(Color online) Extrapolation scheme based on Turnbull's rule: ${\sigma_0 = C_T \Delta  H_f/N_A = C_T T_M\Delta S_f/N_A}$.}
\end{center}
\end{figure}

\begin{figure*}[ht]
\begin{center}
\subfigure {\includegraphics[width=0.48\textwidth,angle=0]{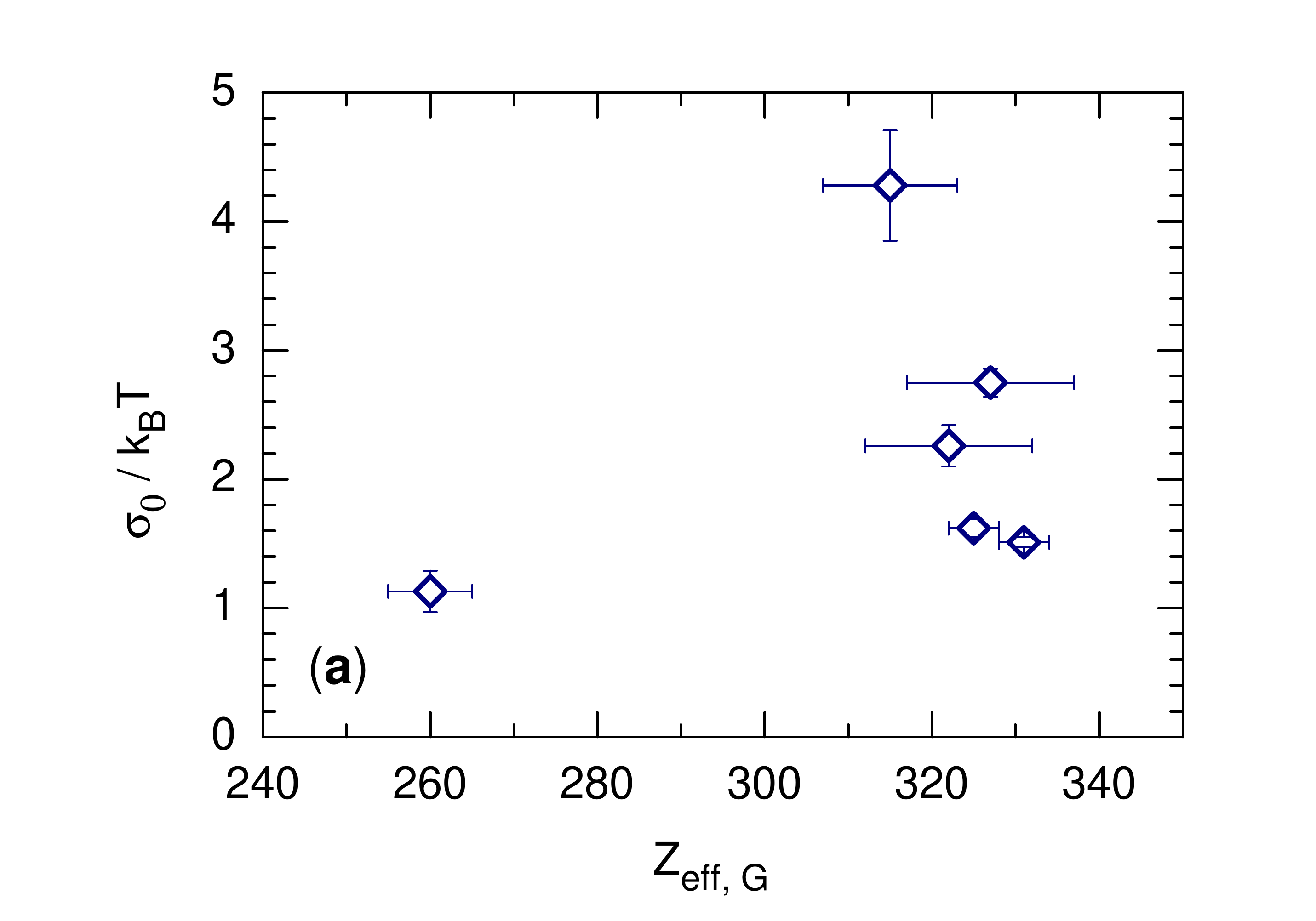}}
\subfigure {\includegraphics[width=0.48\textwidth,angle=0]{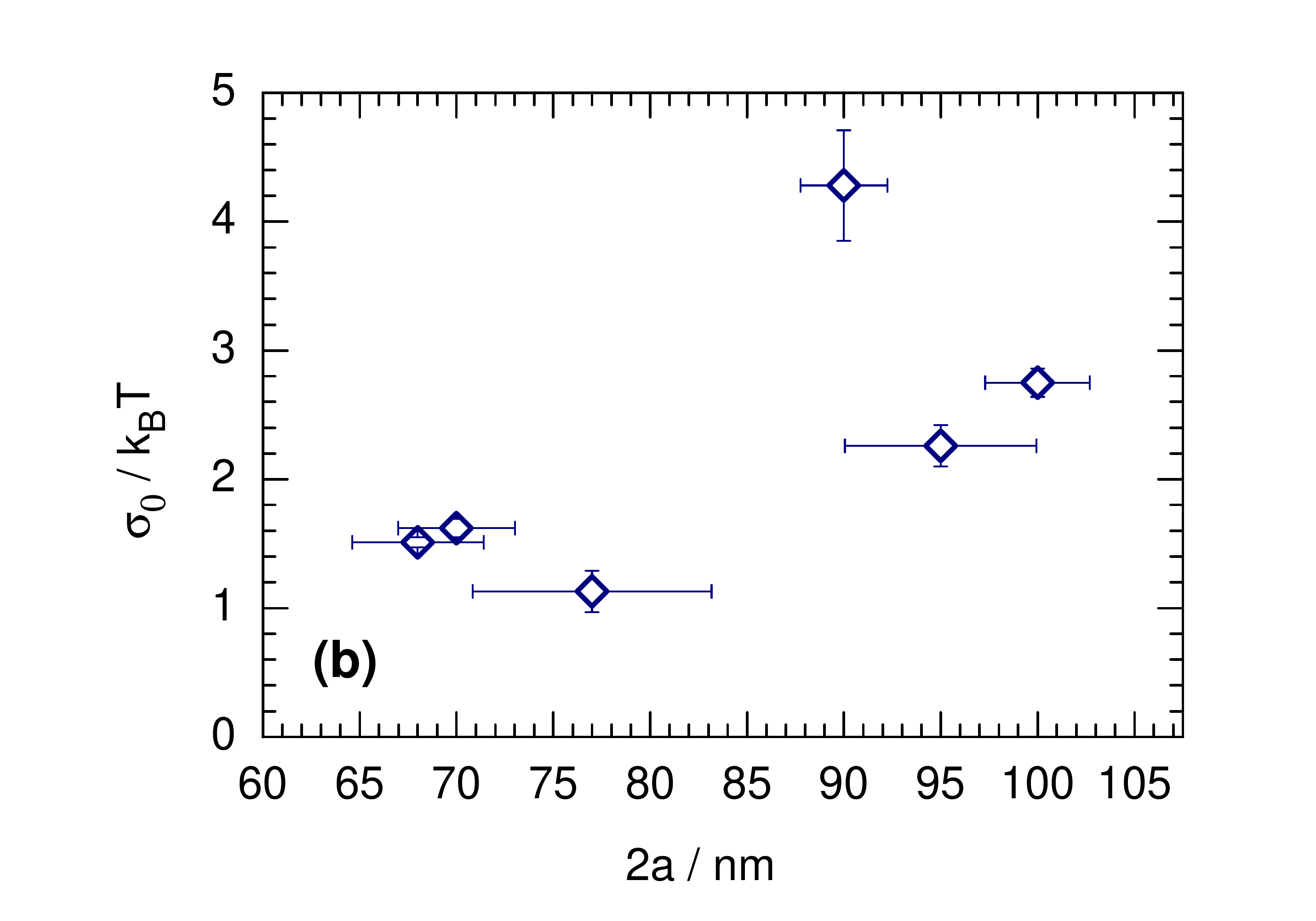}}
\end{center}
\begin{center}
\subfigure {\includegraphics[width=0.48\textwidth,angle=0]{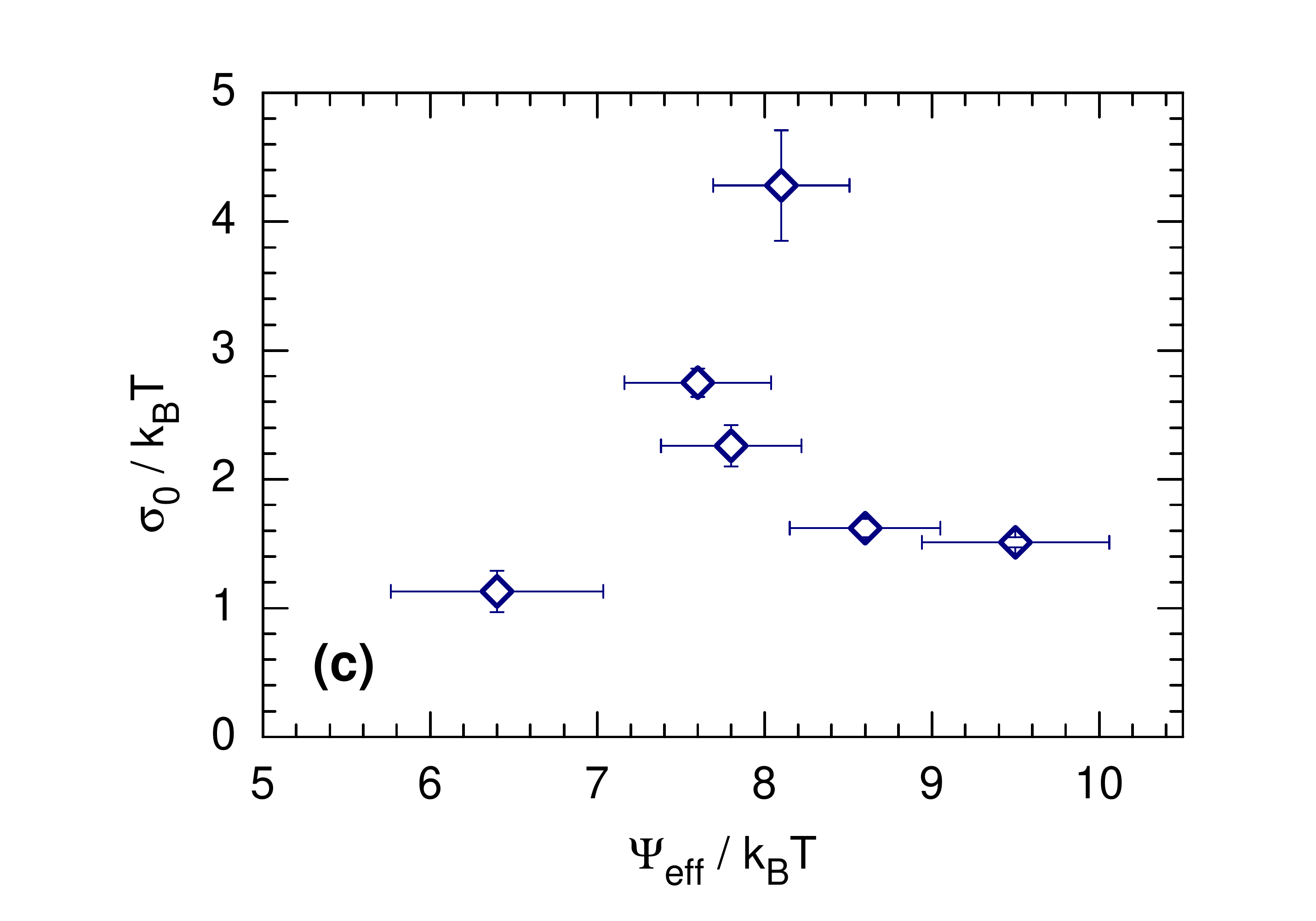}}
\subfigure {\includegraphics[width=0.48\textwidth,angle=0]{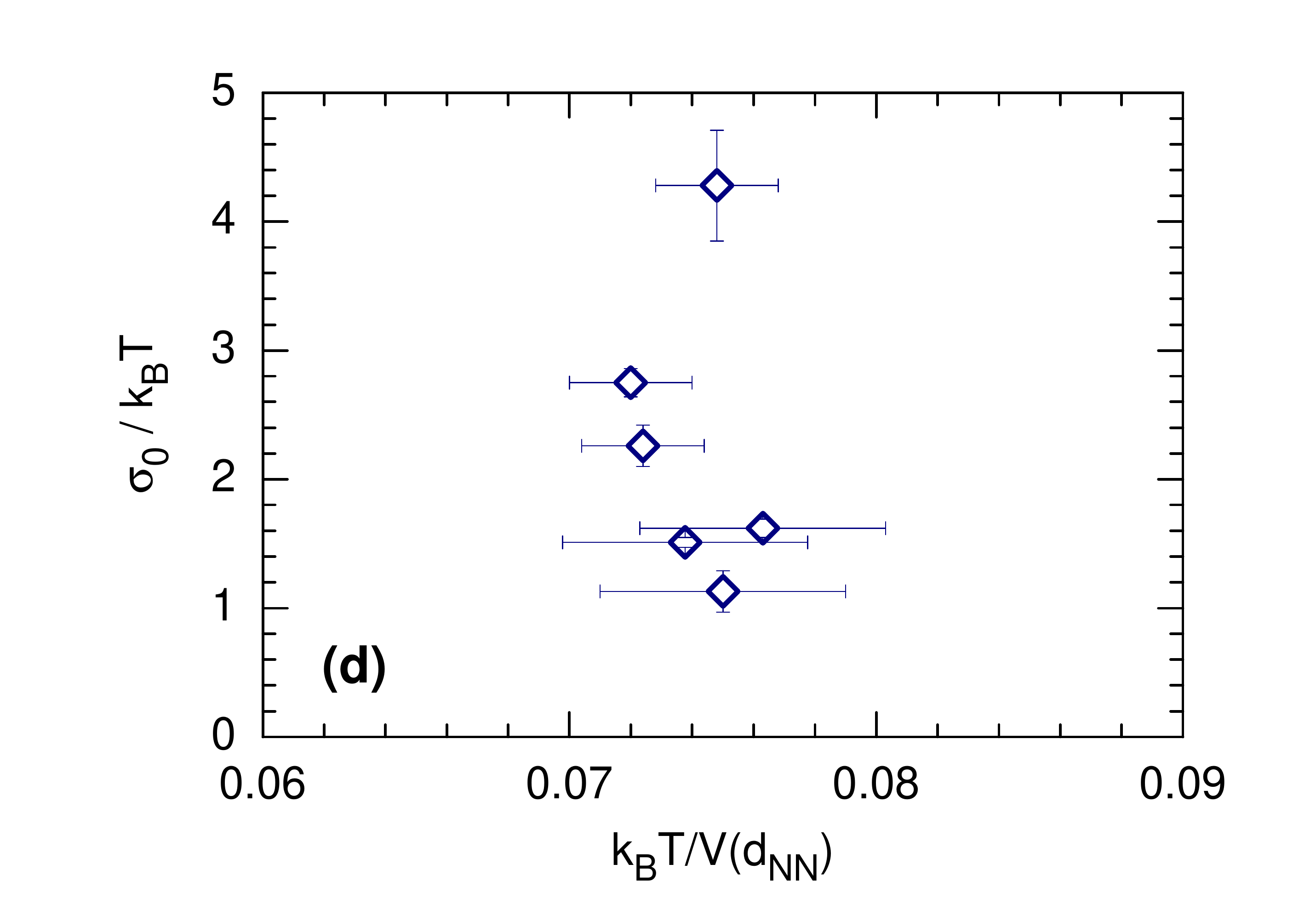}}
\end{center}
\begin{center}
\subfigure {\includegraphics[width=0.48\textwidth,angle=0]{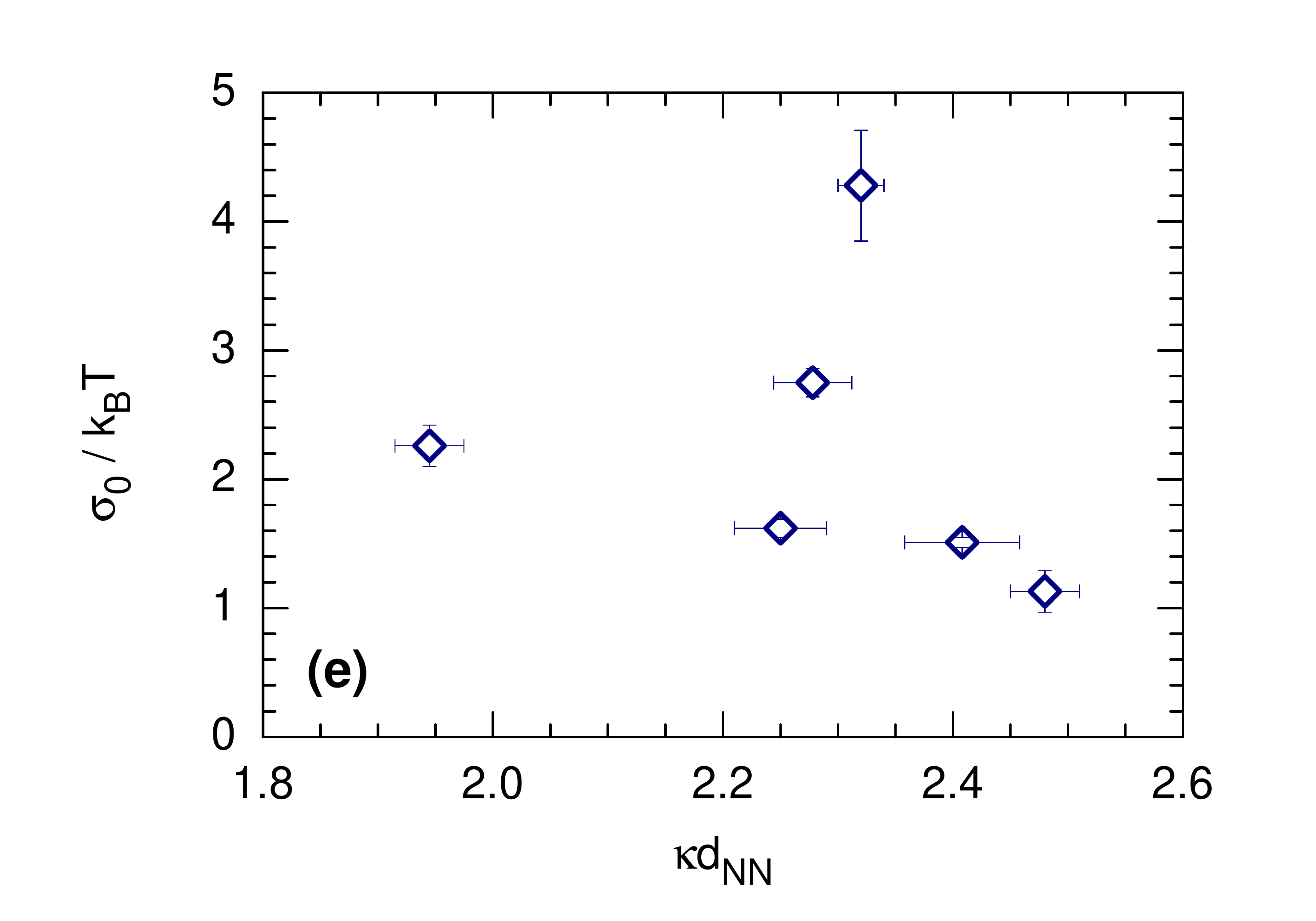}}
\caption{\label{fig3}(Color online) Correlations between the extrapolated reduced equilibrium IFEs and various particle characteristics. (a) effective charges, $Z_{eff,G}$; (b) the number averaged mean particle diameter, 2a; (c) effective surface potential, $\Psi_{eff}$; (d) effective temperature, ${T_{eff}=k_BT/V(d_{NN})}$, at melting, (e) coupling strength at melting; No clear correlation of $\sigma_0$ to any of these interaction parameters is obtained}
\end{center}
\end{figure*}

\begin{figure}[htb]
\begin{center}
\includegraphics[width=0.5\textwidth,angle=0]{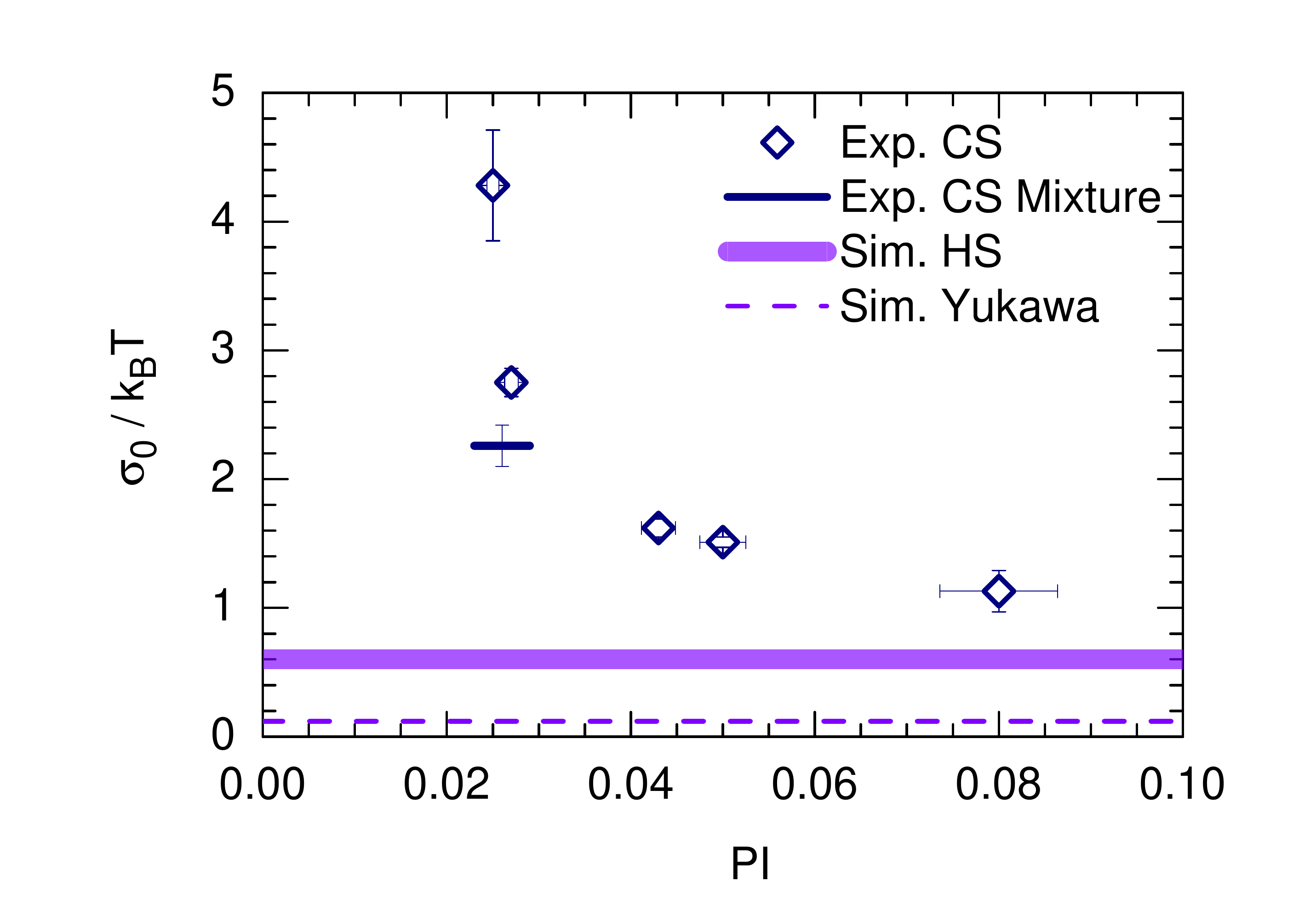}\hspace{0.1cm}
\caption{\label{fig4}(Color online) Correlation between the extrapolated reduced equilibrium IFEs and the polydispersity index PI. A clear anti-correlation is observed: $\sigma_0$ decreases with increasing PI for all single component samples. The case of the mixture is denoted by the horizontal bar. Its length and position denote the range of PIs covered by the involved single component samples. Note, that its position does not correspond to the effective PI of the bimodal size distribution of the mixture. This quantity would be much larger, but cannot be obtained from the equation for monomodal PIs: $PI = s_a/\bar{a}$. Note further, that its value lies significantly below those of the corresponding single component species. For comparison, we also show values for the equilibrium IFE as obtained from simulations of macroscopic flat fluid-crystal interfaces of monodisperse HS (solid horizontal line with thickness corresponding to the spread of published data) and a point Yukawa system (dashed line). (For further details, see text)}
\end{center}
\end{figure}

Note that the obtained $\sigma_0$ range between $1.13k_BT$ and $4.23k_BT$, i.e. a large spread of values is obtained for the different samples. With a thorough system characterization at hand, we check correlations of the obtained $\sigma_0$ to other quantities. In Fig. 3 a-e we plot the data versus a) the effective charges, $Z_{eff,G}$, obtained from elasticity measurements for the crystal phase over the range of interest; b) the number-averaged mean particle diameter, $2a$; c) the dimensionless effective surface potential, $\Psi_{eff}$; d) the effective temperature, ${T_{eff} = k_BT/V(d_{NN})}$, as calculated using $T = 298K$ and the effective interaction strength at the nearest neighbour site (c.f. Eqn.~(B.1) and (B.2)) for the conditions encountered at melting, and e) the value of the coupling parameter ${\lambda=\kappa d_{NN}}$ at melting. No clear correlation is observable in any of these cases.

Next we plot $\sigma_0$ versus the polydispersity index, PI, in Fig.~4. Here, a clear decrease of $\sigma_ 0$ with increasing polydispersity is observed. Further, PS90 and PS100 show $\sigma_0$ of $(4.28 \pm 0.43)k_BT$ and $(2.75 \pm 0.11)k_BT$, respectively. If mixed 1:1 by number, $\sigma_0$ of the PS90/PS100B mixture drops to $(2.26 \pm 0.16)k_BT$ which is significantly lower than the value observed for either pure system. Fig. 4 thus shows that the equilibrium values of the reduced IFE as extrapolated from the CNT-based effective non-equilibrium IFEs are anti-correlated to the polydispersity of the investigated systems. This is a central result of the present paper.

As shown in Appendix A, the slope, $m$, behaves differently and shows neither a correlation to any of the interaction parameters nor to the polydispersity index

\subsection{Estimates for $C_{T,bcc}, \;\Delta  H_f \; \textrm{and} \Delta S_f$}
Our scheme can also be used to extract estimates of further important quantities by making additional assumptions. First, we assume that the molar entropy of freezing does not change with increasing meta-stability, i.e. with increasing particle number density. This has been shown to apply for HS \cite{Laird JCP 2001} and further has been observed for many metal systems \cite{Jiang rev Surf Sci Rep 2008}. Both are systems  where the hard-core repulsion creates an excluded volume which is dominating the behaviour of the condensed phase. In the present systems with their electrostatic interaction, the interaction is much softer, but still the repulsive part dominates the observed ordering processes. Second, we assume that Turnbull's rule which was found for metals also applies in the colloidal case: ${\sigma = C_T \Delta H_f/ N_A}$, where $N_A$ is Avogadro's number. With these assumptions made, we further note that at $\Delta\mu=0$, $\Delta H_f / N_A = T_M \Delta S_f/ N_A$. With ${\Delta S_f/ N_A = const}$ and ${\sigma = C_T \Delta H_f/ N_A}$ this implies that at ${\sigma = 0}$,  ${\Delta H_f/ N_A=0}$ and ${\Delta\mu= -T_M \Delta S_f/ N_A}$, where the melting temperature in our systems is identified with the ambient temperature $T_M = 298K$.  Therefore, extrapolating the data to the intercept with the $\Delta\mu$ axis yields an estimate of the entropy of fusion and the enthalpy of fusion at equilibrium. Finally, the slope of the curve ${m= \sigma_0 / (T_M \Delta S_f/ N_A)}$ can be identified to Turnbull's coefficient $m=C_T$. This latter identification was originally suggested by P.W. in his PhD-thesis \cite{Wette PhD} and was later used in \cite{Herlach JPCM 2011 colloids as models, Herlach Rev EPJST 2014}. Values for $ m = C_{T,bcc}$ are shown in Tab.~II. They range between 0.235 and 0.405, each with small statistical uncertainties reflecting the good linear correlation of $\sigma$ and $\Delta \mu$. The spread of values is smaller that that observed for $\sigma_ 0$ and no clear correlation between $\sigma_ 0$ and $C_{T,bcc}$ is found (see Fig.~7a in Appendix A). Moreover, none of the tests for correlations of $C_{T,bcc}$ to particle or system quantities gave any significant results (see Fig.~8 a-f in Appendix A). In particular $C_{T,bcc}$ is observed to be uncorrelated to the PI.

The values of molar $\Delta S_f$, and $\Delta H_f$ are also compiled in Tab.~II. Those for $\Delta S_f$ range between 1.5 and 4.6 ${J mol^{-1} K^{-1}}$ and those for $\Delta H_f$ range between 0.45 and 1.36 $kJ mol^{-1}$. Correlation checks show that, like $\sigma$, $T\Delta S_f$ (resp. $\Delta H_f$) is not correlated to $Z_{eff}$, $2a$, $\Psi_{eff}$, $T_{eff}$, or $\kappa d_{NN}$. By contrast, an anti-correlation is observed to the PI. Like $\sigma_0$, $T_M\Delta S_f$ shows a clear trend to decreases with increasing PI and again the value for the mixture is well below that of the pure components (see Fig.~7b in Appendix A). To highlight this finding, we plot in Fig.~5 the correlation of $\sigma_0$  to $T_M\Delta S_f = \Delta H_f$ and observe a clear linear correlation with slope $b = 2.76 \pm 0.19$ and a correlation coefficient of $r = 0.988$.

\begin{figure}[htb]
\begin{center}
\includegraphics[width=0.50\textwidth,angle=0]{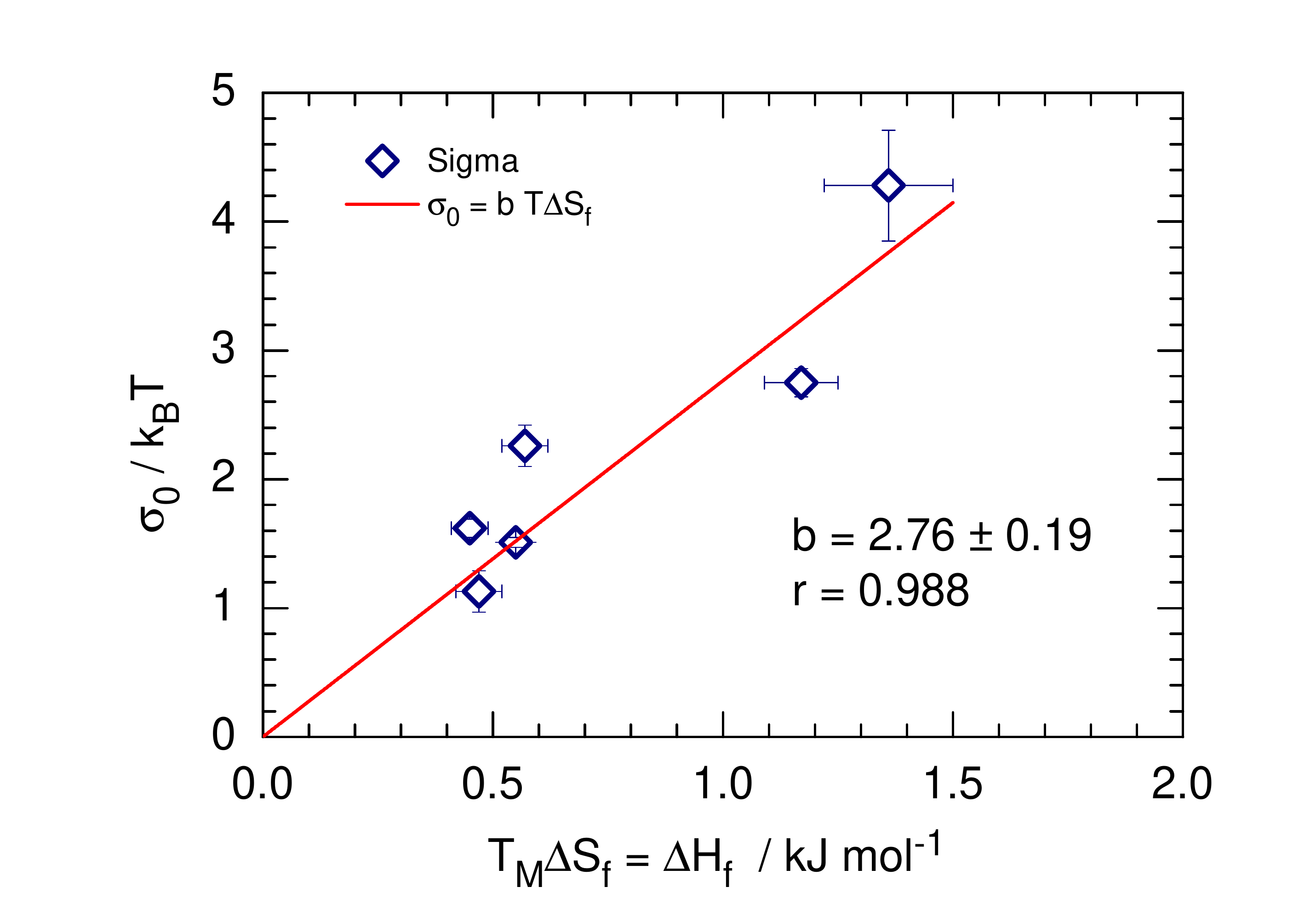}\hspace{0.1cm}
\caption{\label{fig5}(Color online) Correlation between the extrapolated equilibrium reduced IFE and the enthalpy of fusion equalling the entropy of fusion times the melting temperature. A good linear correlation (correlation coefficient r = 0.988) is observed as expected from Turnbull's rule and its interpretation by Laird \cite{Turnbull JCP 1949, Laird JCP 2001}}
\end{center}
\end{figure}

To summarize the results of our extended analysis of CNT-based reduced non-equilibrium effective IFEs, we find that deionized charged sphere suspensions of different charge, size, and polydispersity show extrapolated equilibrium reduced IFEs which are

\begin{itemize} \item[(i)] in the range of a few $k_BT$, \item[(ii)] systematically increasing with increased meta-stability expressed as $\Delta \mu$, \item[(iii)] not correlated to any of the interaction parameters but \item[(iv)] linearly correlated with the entropy of fusion and, \item[(v)]as the latter, decreasing with increasing polydispersity.
\end{itemize}

\section{Discussion}
The above results call for further discussion of a number of points. We first discuss the observed values and their spread, then turn to the observed anti-correlation of $\sigma_0$ to the PI and finally comment on the observed Turnbull coefficients for bcc crystallizing systems.

The magnitude of the observed equilibrium reduced IFEs may appear unexpectedly large. Values range between 1.25kBT and 4.4kBT. In Fig. 4, we compare it to values observed for other systems. The lowest values reported so far were obtained for the point-Yukawa system by Heinonen et al. \cite{Heinonen JCP 2014} as $\sigma_{0,bcc,\textrm{Yukawa}} = (0.12 \pm 0,02)k_BT$ (dashed line in Fig.~4). These authors further mention unpublished work, where they obtained ${\sigma_{0,fcc,Yukawa} = 0.4k_BT}$ (Ref. [46] in \cite{Heinonen JCP 2014}), whereas their value obtained using the same methods for HS was ${\sigma_{0,fcc,HS} = 0.65k_BT}$. Auer and Frenkel have given CNT-based estimates of non-equilibrium IFEs from their Monte Carlo simulations of slightly charged hard spheres modelled by a repulsive hard-core potential \cite{Auer JPCM 2002 weakly charged, Auer AnnRevPC 2004} which are ${\gamma (2a)^2 = 0.45k_BT}$ in the case of fcc and ${\gamma (2a)^2 = 0.38k_BT}$ in the case of bcc crystals. Normalization with ${{d_{NN}^2}}$ would have resulted in somewhat larger values for the reduced non-equilibrium IFEs. However, given the experimentally observed dependence of $\sigma$ on meta-stability, one would expect the equilibrium values to be much smaller than the reported non-equilibrium values, possibly close to those found by \cite{Heinonen JCP 2014}. In any case, these values are considerably smaller than values reported for HS and the extrapolated IFEs reported here.

In Fig. 4, we also display the range of values reported for the HS reference system, $\sigma_{0,fcc,HS} = (0.56-0,68)k_BT$ (thick horizontal line). We note that CNT-based effective IFEs \cite{harland PRE 55, Palberg JPCM Rev 1999, Auer Nature 2001 HS mono, Auer AnnRevPC 2004, Gasser JPCM 2009, Palberg JCPM Rev 2014, Franke SM 2014} and equilibrium IFEs \cite{Hartel PRL 2012 HS, zykova-Timan JCP 2010 HS AHS, davidchack JPCB 2005 MD HS, Mu JPCB 2005 anisotropy HS, davidchack JCP 2006 fluct HS, Amini PRB 2008 fluct HS, davidchack JCP 2010 HS,  Fenandez PRL 2012 HS, Hernández-Guzmán PNAS 2009, Rogers Phil Mag 2011} do not show significant differences and are all significantly below the values observed here for the case of CS.

Metals on the other hand show values of ${20-400 mJ/m^2}$ at their nucleation temperature. When scaled to $T_M$ and the area taken by a single atom in the interface, this results in values of several to some tens of $k_BT$ for $\sigma_{0,fcc,metal}$ \cite{Turnbull JCP 1949, Turnbull JAP 1950, Antonovicz Rev Adv Mat Sci 2008, Kelton Solid state Phys 1993, Jiang rev Surf Sci Rep 2008,  MRS 29 2004, Hoyt MRS 2004}. Of particular interest is the recent comparison of nucleation barriers of Ni derived \textit{via} CNT from nucleation experiments to those derived from state of the art Monte Carlo simulations on the nucleating system \cite{bokeloh PRL 2015} which quantitatively coincide in the case of sufficiently large simulated systems. Further, a mere 10\% discrepancy is observed between the derived non-equilibrium IFEs scaled to the melting temperature and the equilibrium values derived from simulations of the equilibrated interface \cite{Rozas EPL 2011}.

A good agreement is thus observed between CNT-based effective IFEs and the more directly measured equilibrium IFEs for HS and metals, while the present CNT-based effective IFEs of polydisperse CS differ from the equilibrium IFEs obtained for Yukawa and hard-core Yukawa systems. Thus either CNT is not very reliable in our case or mean field descriptions are not suitable to predict IFEs for the systems used in our experiments, even though Yukawa and hard-core Yukawa potentials quantitatively capture the interaction strength and range,.

Our comparison reveals a pronounced sorting in groups of different interaction type. In principle, also the significant spread within the group of charged colloids (and that of metals) can be due to differences in interaction type, strength and/or range. However, as seen in Fig. 3a-e, there is no significant correlation between $\sigma_ 0$ and any of the interaction parameters. On the other hand, we observed a clear anti-correlation between $\sigma_0$ and the PI in Fig.~4 and a linear correlation of $\sigma_0$ to $T_M \Delta S_f$ in Fig.~5. It should be noted that already in his seminal paper \cite{Turnbull JAP 1950}, Turnbull observed $\sigma_ 0$ to be correlated to the melting temperature $T_M$ of the investigated elements. However, due to the rather large overall scatter of this data, this observation was discarded as basis to formulate an empirical relation. Rather, the clearer correlation to $ \Delta H_f/N_A$ was used. Later, Laird theoretically investigated the IFE of hard-core systems to observe a clear scaling of the metal $\sigma_ 0$ with $T_M$ \cite{Laird JCP 2001}. This correlation is also present in the much larger data compilation published by \cite{Jiang rev Surf Sci Rep 2008}. Laird pointed out that this scaling is a direct consequence of the purely entropic determination of the phase behavior of HS and the presence of a hard-core like repulsion in metals. Earlier, Saepen et al \cite{Spaepen Acta Metallica 1975, Spaepen Scr Metallica 1976} similarly argued that given a structure specific but otherwise constant entropy of fusion, the IFE should vary linearly with temperature upon undercooling a melt. This is also seen for the present systems, where $\sigma_ 0$ varies linearly with ${\Delta \mu}$ which may be regarded as the colloid analog of undercooling. Therefore, we believe that the presently observed large values for $\sigma_0$ are caused by a large entropy difference between melt and crystal and furthermore, that the spread is caused by a polydispersity-induced variation of this difference.

We may rationalize this, considering a monodisperse system transforming from a melt of short range order to a crystalline state of long range order. Introducing some polydispersity will disturb both phases differently. There will be a structural change (a deviation from the best ordered low energy / low entropy configuration) which is different for the polydisperse crystal and for the polydisperse melt. Loosely speaking, the order of the melt will be disturbed only over short distances, while the disturbance in the arrangement of particles in the crystalline state will be long ranged. In both cases, the entropy of the less well ordered region will increase, but, due to the rules of combinatorics, the effect of the larger number of particles involved in the distorted crystal is far larger. Therefore, in the polydisperse case, the entropy difference between the two phases will be smaller than in the monodisperse case. Consequently, also $\sigma_0$ has to decrease, and we observe the samples with the largest PI to display the lowest IFE. We note that the results shown in Fig. 4 and 5 may be considered as an independent test of Turnbull's rule using model systems of different polydispersity yielding an \textit{a priori} unknown, but systematic decrease of entropy differences.

We come back to the discrepancies observed in comparing our IFEs to those of simulated point and hard-core Yukawa systems. For the presently examined case of experimental charged spheres with their additional $Z_{eff}$ counter-ions per particle, the entropy difference between melt and solid should be much larger than for any HS, point Yukawa or hard-core Yukawa system. In this latter cases, either counter-ions are not present or are absorbed in a neutralizing mean field back-ground. Therefore, any counter-ion contribution to the entropy is neglected in these systems. In principle, this hypothesis can be tested by theoretical investigations within the primitive model that may become feasible with state-of-the-art algorithms \cite{Kratzer SM 2015}. Moreover, with this data available also a comparison of the presently derived CNT-based effective IFEs and more directly measured equilibrium IFEs will become possible.

Crystallization of colloidal suspensions may involve fractionation processes. The PIs of the present systems ($0.025 \leq PI \leq 0.08$) are considerably lower than those for which fractionation is expected and/or found in CS \cite{VanderLinden JCP 2013, Labbez 2015 preprint}. Further, in none of the experiments the characteristic broad, pyramid-shaped Bragg peaks were observed \cite{Kozina SM 2012}. Moreover, in deionized CS binary mixtures, the size ratios for stabilizing azeotropic or eutectic phase behaviour are shifted to much smaller values as compared to HS \cite{Lorenz JPCM 2009}. Therefore, we believe that the effective CS polydispersities are considerably lower than the geometrical ones, and we exclude fractionation effects for the present systems. However, in accordance with theoretical expectations \cite{Fasolo PRL 2003, Fasolo PRE 2004, Sollich PRL 2010, Sollich SM 2011}, fractionation has been observed in strongly polydisperse experimental HS systems \cite{Martin PRE 2003, Kozina SM 2012}.

Fractionation has also been observed in the simulations of Auer and Frenkel \cite{Auer Nature 2001 polydisperse, Auer AnnRevPC 2004}. They reported an increase of the nucleation barrier with increasing PI for $PI > 0.05$ which was attributed to an increase in IFE. This seems to be at odds with the present observation of a clear non-linear decrease of $\sigma_ 0$ with increasing PI (c.f. Fig.~4). However, fractionation affords the formation of purified phases of much lower entropy than the melt or the substitutional crystal \cite{Ganagalla JCP 2013}. For the purely entropic HS system, this will increase both $\gamma$ and $\Delta \mu$ between the fractionated crystal nucleus and the remaining melt. Consequently, the nucleation barrier ${\Delta G^*_{CNT}= (16\pi\gamma^3)/3{(n \Delta\mu)}^2}$ will increase with increasing PI. This effect limits the range of applicability of our conclusions to non-fractionating systems. The  presently observed decrease of $\sigma_ 0$ with PI should be reversed at the onset of fractionation, and the IFE should display a minimum as a function of PI. The situation is further complicated by the fact, that fractionation will also influence the kinetic pre-factor. The required sorting processes decreases the kinetic pre-factor, and contribute to the observed drastic slowing of nucleation \cite{Ganagalla JCP 2013, Kozina SM 2014}. The findings by Sch\"ope \cite{Schope PRE 2006 polyd} may have been made in a cross-over region, where the decreased kinetic pre-factor already caused considerably stretched induction stages, while the effects on the barrier were not yet pronounced enough to quench nucleation effectively. However, the issue of the influence of fractionation is far from being generally settled. It should be addressed again in both simulation and experiments on non-fractionating HS systems as well as fractionating CS systems.

We further compare the values of the other estimated thermodynamic quantities to those for other systems. Our absolute values for molar $\Delta H_f$ range between 0,45 and 1.36 ${kJ mol^{-1}}$. This is just below those of alkaline metals, which range about ${(2-3) kJ mol^{-1}}$, but smaller than the values of about ${8 kJ mol^{-1}}$ for alkaline earth metals and 35.2 $kJ mol^{-1}$ for Tungsten \cite{Tungsten Web}. Values for $\Delta S_f$ range between 1.5 and 4.6 $J mol^{-1} K^{-1}$ as compared to a value of about 10 $J mol^{-1} K^{-1}$ for metals. For monodisperse HS, density functional theory calculations yield 9.7 $J mol^{-1} K^{-1}$ \cite{Laird JCP 2001}. The values for polydisperse CS are thus much smaller than those of monodisperse systems with shorter-ranged interactions. Therefore, it would be interesting to pursue this case further in two ways. One way is to investigate nucleation in charged sphere suspensions at larger particle and salt concentrations, where the interaction potential becomes less soft. The second way would be investigations using CS with vanishing PI.

Finally, we have used the identification of the fitted slope $m$ to Turnbull's coefficient $C_{T,bcc}$. From this, we obtained the first systematic estimates for this quantity for a set of experimental systems crystallizing into bcc structure. We note that as expected, the presently observed values are independent of $\sigma_0$, of the PI and of the interaction parameters (c.f. Appendix A). In fact, the strictly linear dependence of $\sigma$ on $\Delta\mu$ shows that the very same Turnbull coefficients also apply in the meta-stable state. Up to now, experimental determinations have only been performed for fcc crystallizing metals, yielding $C_{T,fcc} = 0.43$ \cite{Kelton Solid state Phys 1993}, while simulation results for fcc crystallizing metals are better described by $C_{T,fcc} = 0.55$ \cite{Hoyt MRS 2004}. The few simulations available for bcc crystallizing metals are best described by $C_{T,bcc} = 0.29$ \cite{Hoyt MRS 2004}. In Fig.~6 we plot the obtained $C_{T,bcc}$ versus the corresponding reduced equilibrium IFEs (in eV per atom) against the equilibrium enthalpy of fusion (in eV per atom). From an error-weighted linear fit to the data of Fig.~6, we find an averaged Turnbull coefficient of $C_{T,bcc,exp} = 0.31\pm 0.03$. Using only the low-uncertainty data with $\Delta \mu$ derived using W\"urth's approximation for $\Delta\mu$, we obtain a slightly lower value of ${C_{T,bcc,exp} = 0.25\pm0.02}$. For comparison, we also show the currently available data from the literature. Our values appear to be remarkably close to those from simulations of bcc metals (open triangles \cite{Hoyt MRS 2004}) which yielded an average ${C_{T,bcc,sim} = 0.29}$ (dash-dotted line) but are much smaller than the values for fcc crystallizing systems. This further supports the theoretical expectations based on entropic considerations that predict that $C_{T,bcc}$ should be considerably smaller than $C_{T,fcc}$ \cite{Spaepen Acta Metallica 1975, Spaepen Scr Metallica 1976, Granazy MDF 1991 broken bond model}.

 \begin{figure}[htb]
\begin{center}
\includegraphics[width=0.50\textwidth,angle=0]{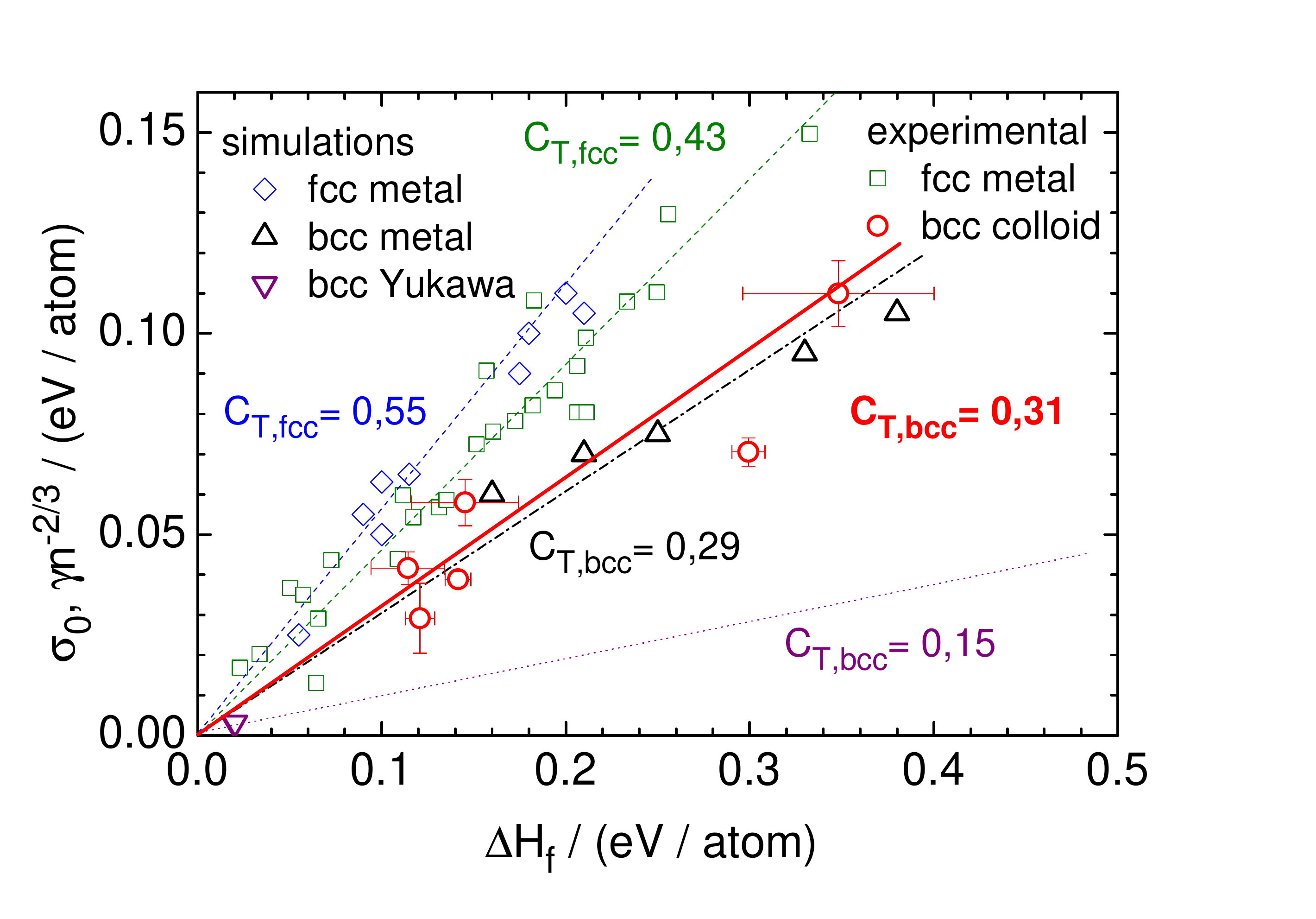}\hspace{0.2cm}
\caption{\label{fig6}(Color online) Turnbull plot of the reduced equilibrium interfacial free energy versus the equilibrium enthalpy of fusion. Shown are our data for bcc crystallizing colloids (circles), simulation data for bcc crystallizing metals (up triangles) \cite{Hoyt MRS 2004}, bcc crystallizing point Yukawa systems (down triangles) \cite{Heinonen JCP 2014} and fcc crystallizing metals (diamonds) \cite{Hoyt MRS 2004} as well as experimental data for fcc crystallizing metals (squares) \cite{Kelton Solid state Phys 1993}. Lines correspond to the indicated average values of Turnbull coefficients as quoted from \cite{Hoyt MRS 2004,Heinonen JCP 2014}, and \cite{Kelton Solid state Phys 1993} for simulation and experimental data, respectively. For our data we find an average value of $C_{T,bcc} = 0.31\pm0.03$ (thick solid line) which appears to be remarkably close to that expected for bcc metals}
\end{center}
\end{figure}

\section{Conclusions}
We have analyzed data available from the literature on five pure species and one mixture of charged colloidal spheres obtained from crystallization experiments under deionized conditions. We have devised a simple extrapolation scheme to obtain estimates for the reduced equilibrium IFE from the reduced non-equilibrium CNT-based effective IFEs. Under the additional assumptions of a system specific entropy of fusion which, however, is independent of system density and the validity of Turnbull's rule, we further used this scheme to extract the enthalpy and entropy of fusion, $\Delta H_f$ and $\Delta S_f$, as well as the Turnbull coefficient $C_{T,bcc}$. The latter data complements existing experimental and theoretical data on fcc crystallizing systems. The experimental $C_{T,bcc}$ was found to be remarkably close to expectations from simulations on bcc crystallizing metals. This strongly supports the results of computer simulation and theoretical models predicting a substantial difference in $C_T$ for differing crystal structures.

Incidentally, all analyzed experimental samples showed different degrees of polydispersity. This allowed a discussion of the influence of the PI on both the interfacial free energy and the entropy of freezing. We observed that both quantities show a similar clear trend to decrease with increasing PI and that both quantities are linearly correlated to each other. From this, we conclude that an increase in polydispersity lowers the crystal-melt entropy difference and - consequently - the IFE. Thus, we believe that the thermodynamics of freezing in charged sphere suspensions are dominated by entropic rather than enthalpic effects. At least for the present system, we suggest to reformulate Turnbull's rule in terms of the melting temperature times the gram-atomic entropy of fusion.

The values of the extrapolated IFEs range between those of metal systems and of hard spheres. They show a clear tend to increase with decreasing PI. A PI-dependent IFE has interesting possible consequences for the kinetics of nucleation. If the increase in $\sigma_ 0$ continued for still lower PI, one would expect the nucleation barriers to become very large in the limit of monodisperse charged sphere systems. Then crystallization might become suppressed in favor of a Wigner-glass \cite{Yazdi PRE 2014 Wigner Glass}. On the other hand, it may be beneficial to carefully re-investigate the influence of a small geometric polydispersity (for example PI $\leq 0.03$) on the nucleation barrier of HS systems. There, state-of-the-art simulations on monodisperse systems in the coexistence region yield nucleation rate densities which consistently are several orders of magnitude lower than those observed in polydisperse experimental systems \cite{Palberg JCPM Rev 2014}. If HS showed a similar dependence of $\sigma_ 0$ on PI as charged spheres, one would expect a small polydispersity to considerably accelerate nucleation.

In most previous work on experimental and computer hard spheres as well as on metals, a near quantitative agreement between CNT-based effective IFEs and those measured more directly on equilibrated interfaces was obtained. Therefore, the observed disagreement between the presently derived results for the CNT-based effective equilibrium IFE and the equilibrium IFE obtained for point Yukawa-systems in simulations calls for further attention. Typically, both point Yukawa and hard-core Yukawa based simulations yield good results in predicting charged sphere suspension properties including phase behaviour, elasticity and electro-kinetic behaviour. A hard-core Yukawa potential has also been employed in the evaluation of the crystallization kinetic data used for the present analysis. Therefore, the reason for the observed discrepancy remains unclear. It may be that the parametrization of measured nucleation rate densities using CNT is not appropriate for the case of CS. However, in view of our main finding that the IFE is mainly an entropic effect, we are tempted to ascribe the discrepancy to the use of Yukawa-type mean field pair potentials in the simulations made. Hence, it will be very interesting to compare our data to future simulations within the primitive model with explicit counter-ions which may better capture the entropic contributions of the micro-ions.

\section{Acknowledgements} We are pleased to thank J. Horbach, K. Binder, Th. Speck, Th. Voigtmann, H. L\"owen, and H. J. Sch\"ope for many fruitful, and sometimes critical, discussions. We further thank A. Engelbrecht, R. Meneses, N. Lorenz, I. Klassen and H. J. Sch\"ope for providing access to the original data beyond published results, the referees of Phys. Rev. for valuable suggestions and B. Robertson for reading the manuscript. Financial support by the DFG (Pa459/16,17; He1601/24) is gratefully acknowledged.

\bigskip

\textbf{Appendix A: Additional correlation checks}

Because the data was available, we performed a number of additional checks for correlations between the obtained key parameters of crystallization and the experimental and system specific boundary conditions.
\begin{figure}[htb]
\begin{center}
\includegraphics[width=0.5\textwidth,angle=0]{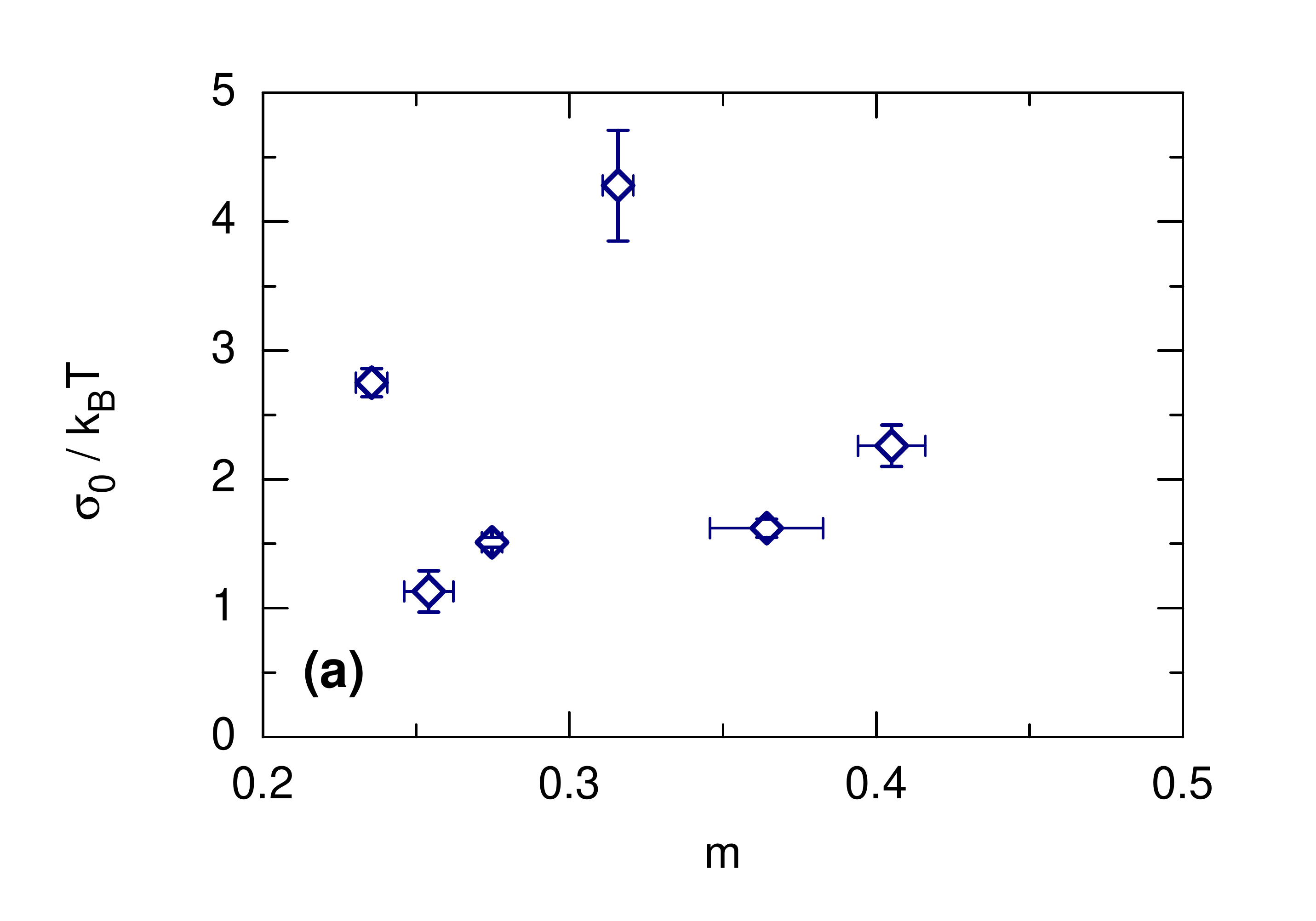}\hspace{0.2cm}
\includegraphics[width=0.5\textwidth,angle=0]{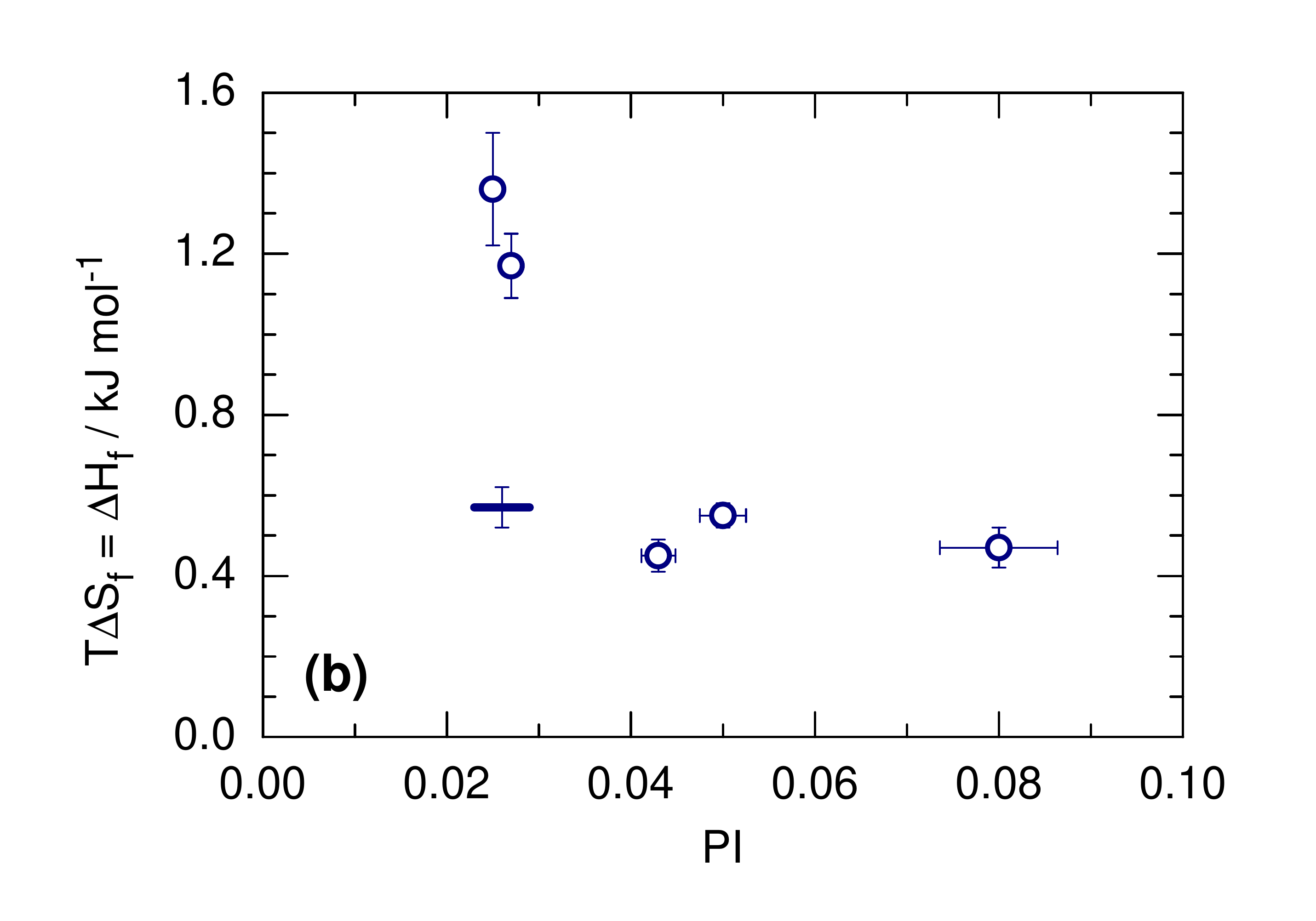}\hspace{0.2cm}
\caption{\label{fig7}(Color online) (a) equilibrium reduced IFEs $\sigma_ 0$ versus $m$. No correlation between these two quantities is observed. (b) plot of $T_M\Delta S_f = \Delta H_f$  versus the polydispersity index, PI. A clear decrease is observed. In addition, note the strong decrease also for the value of the mixture (horizontal bar), as compared to those of the two pure components (leftmost circles).}
\end{center}
\end{figure}

\begin{figure*}[ht]
\begin{center}
\subfigure{\includegraphics[width=0.48\textwidth,angle=0]{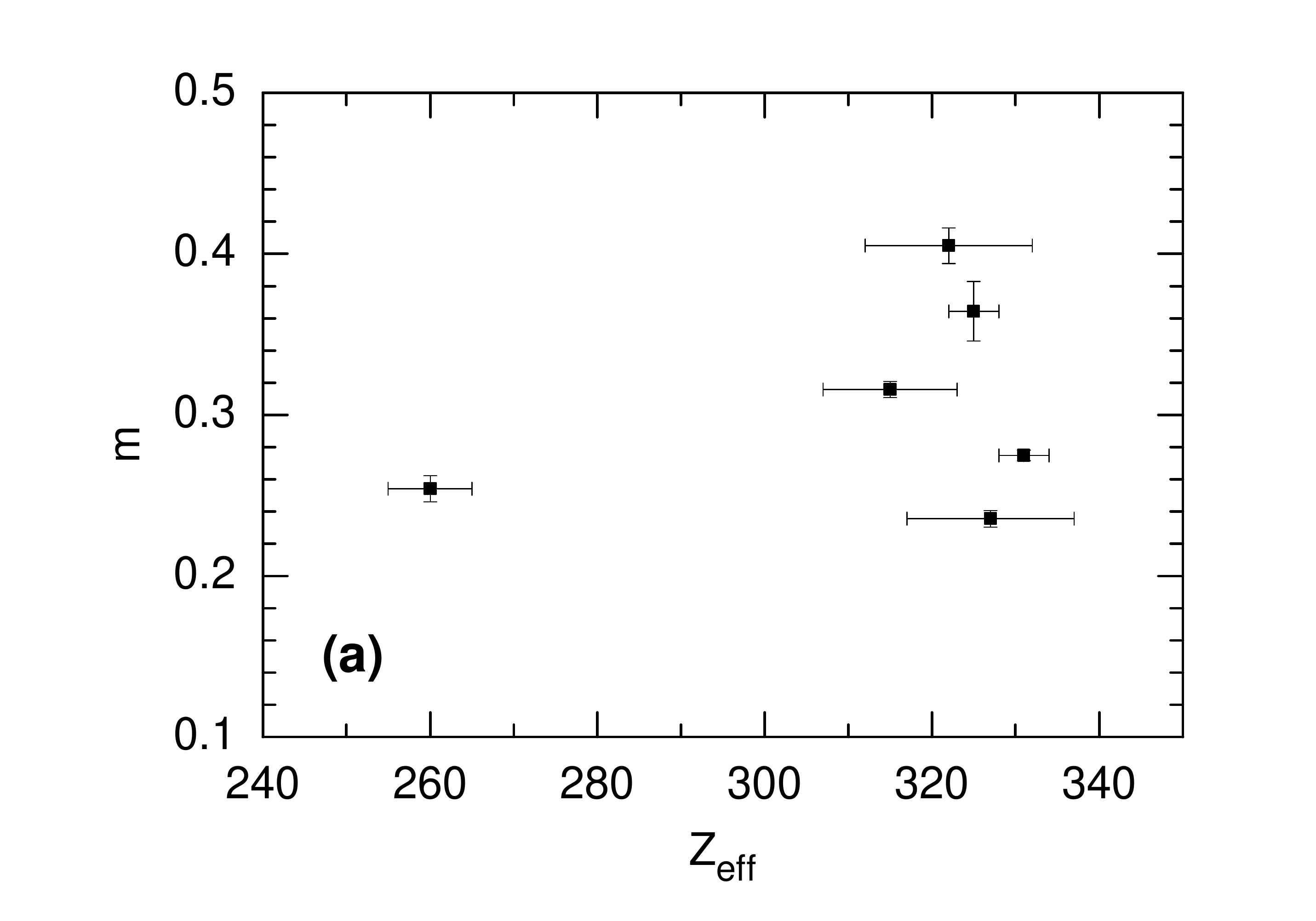}\hspace{0.2cm}}
\subfigure{\includegraphics[width=0.48\textwidth,angle=0]{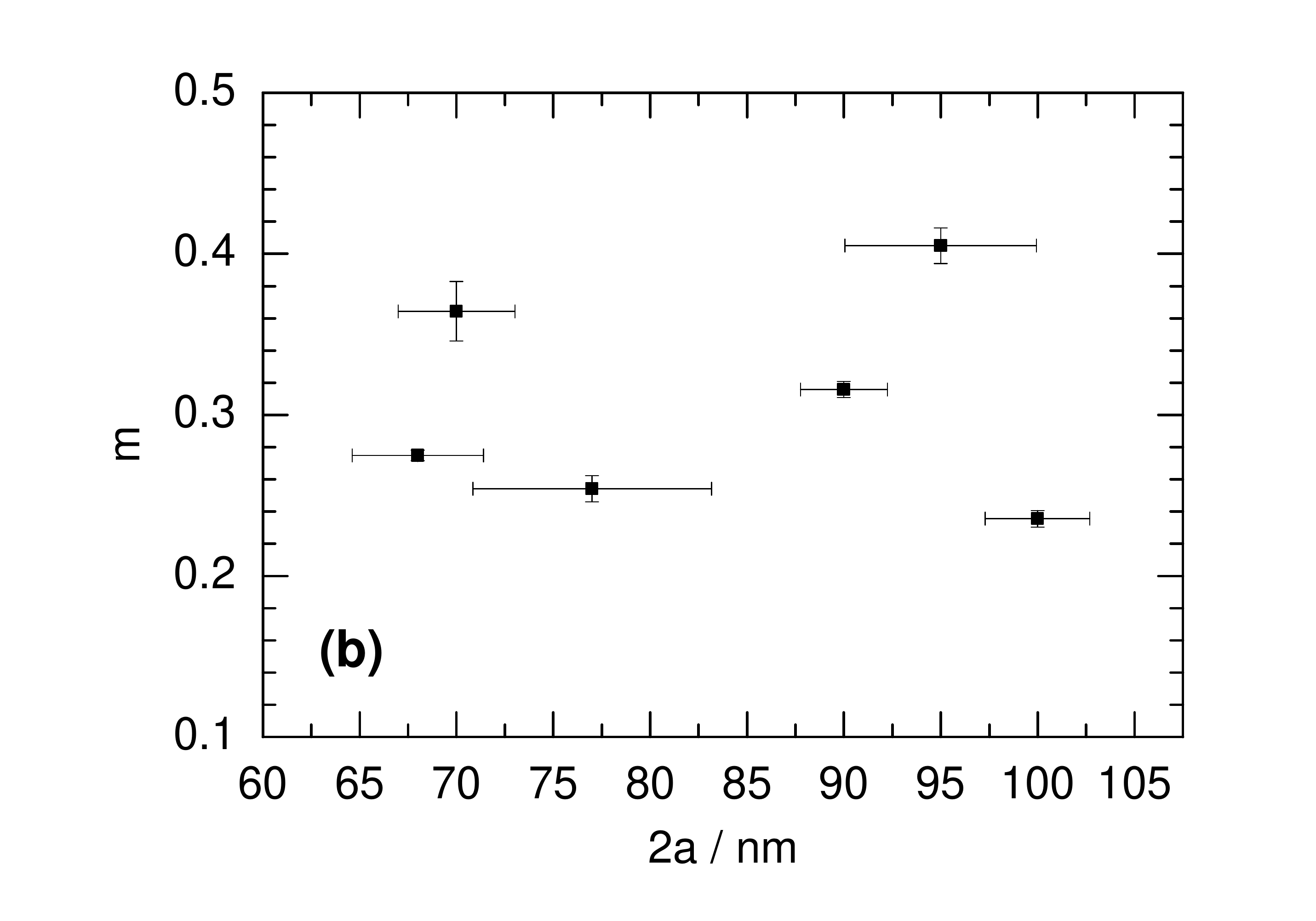}\hspace{0.2cm}}
\end{center}
\begin{center}
\subfigure{\includegraphics[width=0.48\textwidth,angle=0]{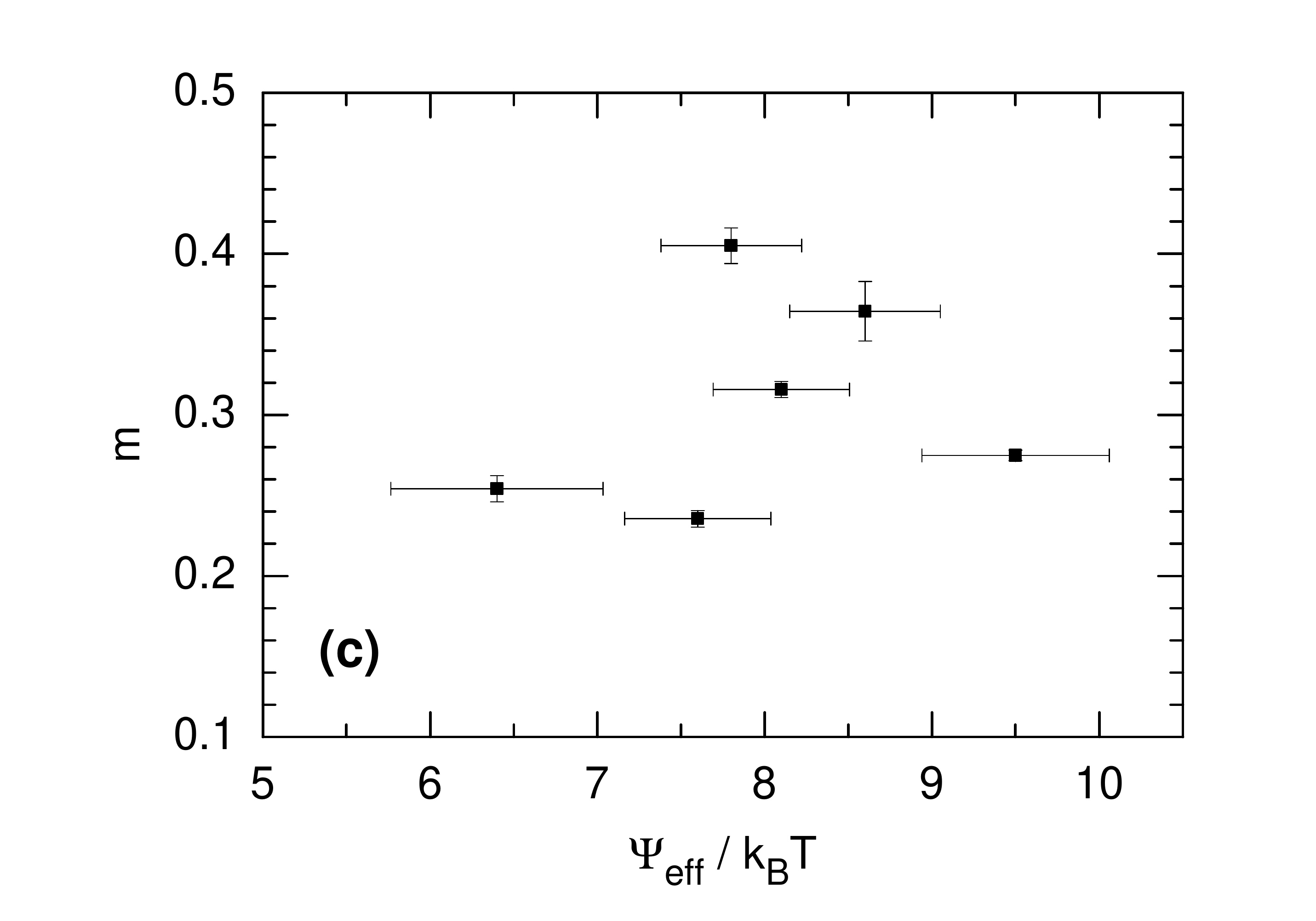}\hspace{0.2cm}}
\subfigure{\includegraphics[width=0.48\textwidth,angle=0]{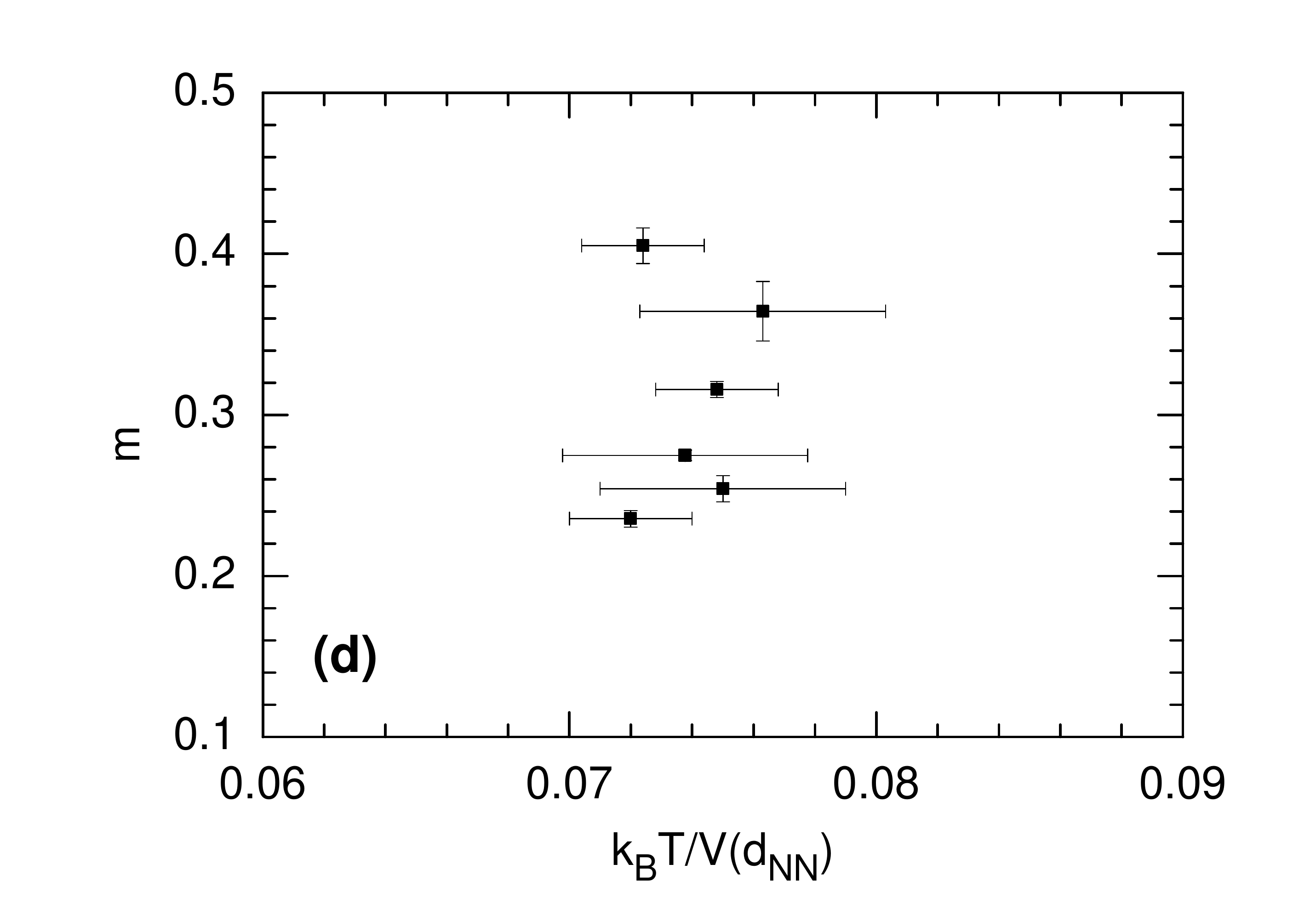}\hspace{0.2cm}}
\end{center}
\begin{center}
\subfigure{\includegraphics[width=0.48\textwidth,angle=0]{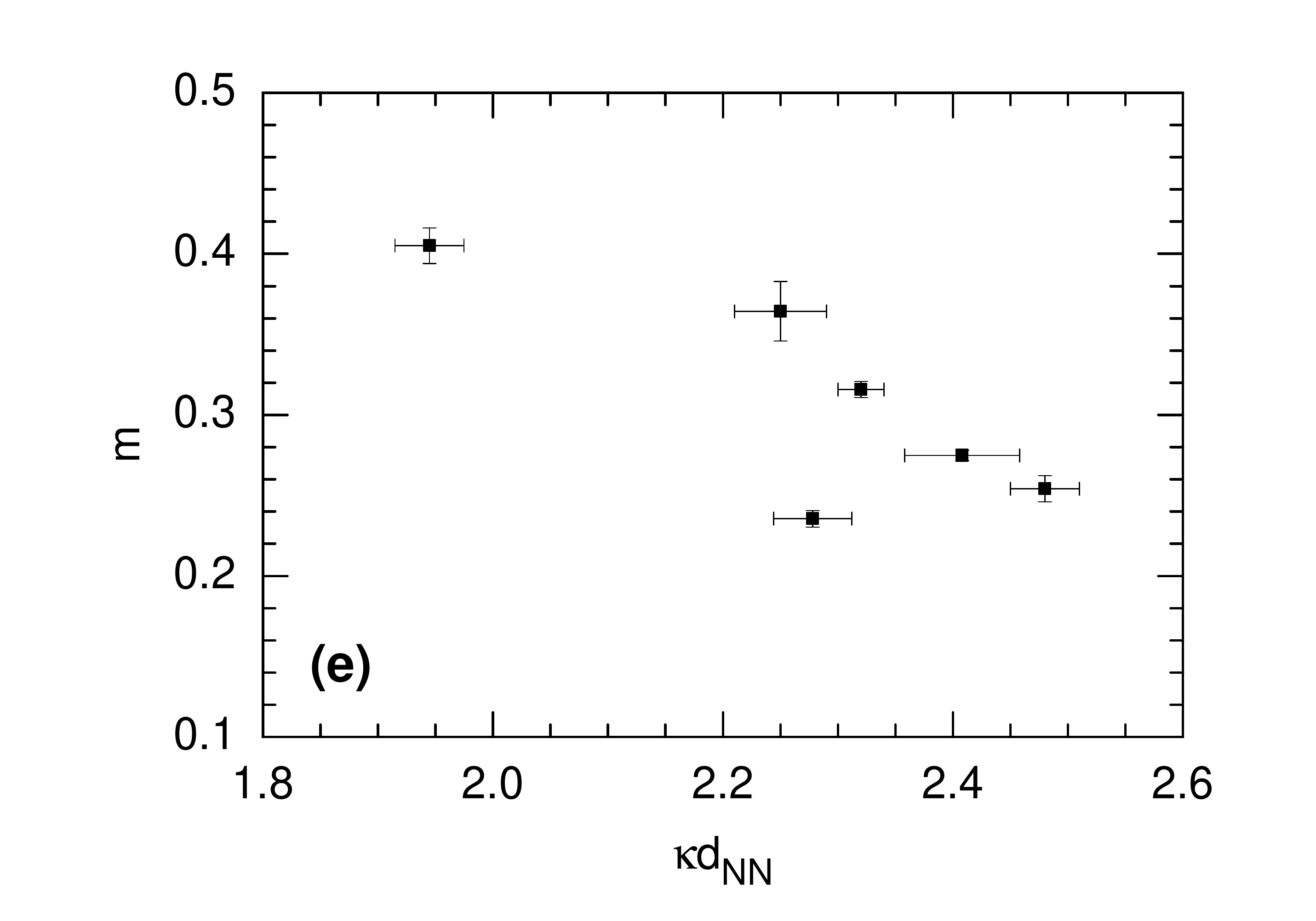}\hspace{0.2cm}}
\subfigure{\includegraphics[width=0.48\textwidth,angle=0]{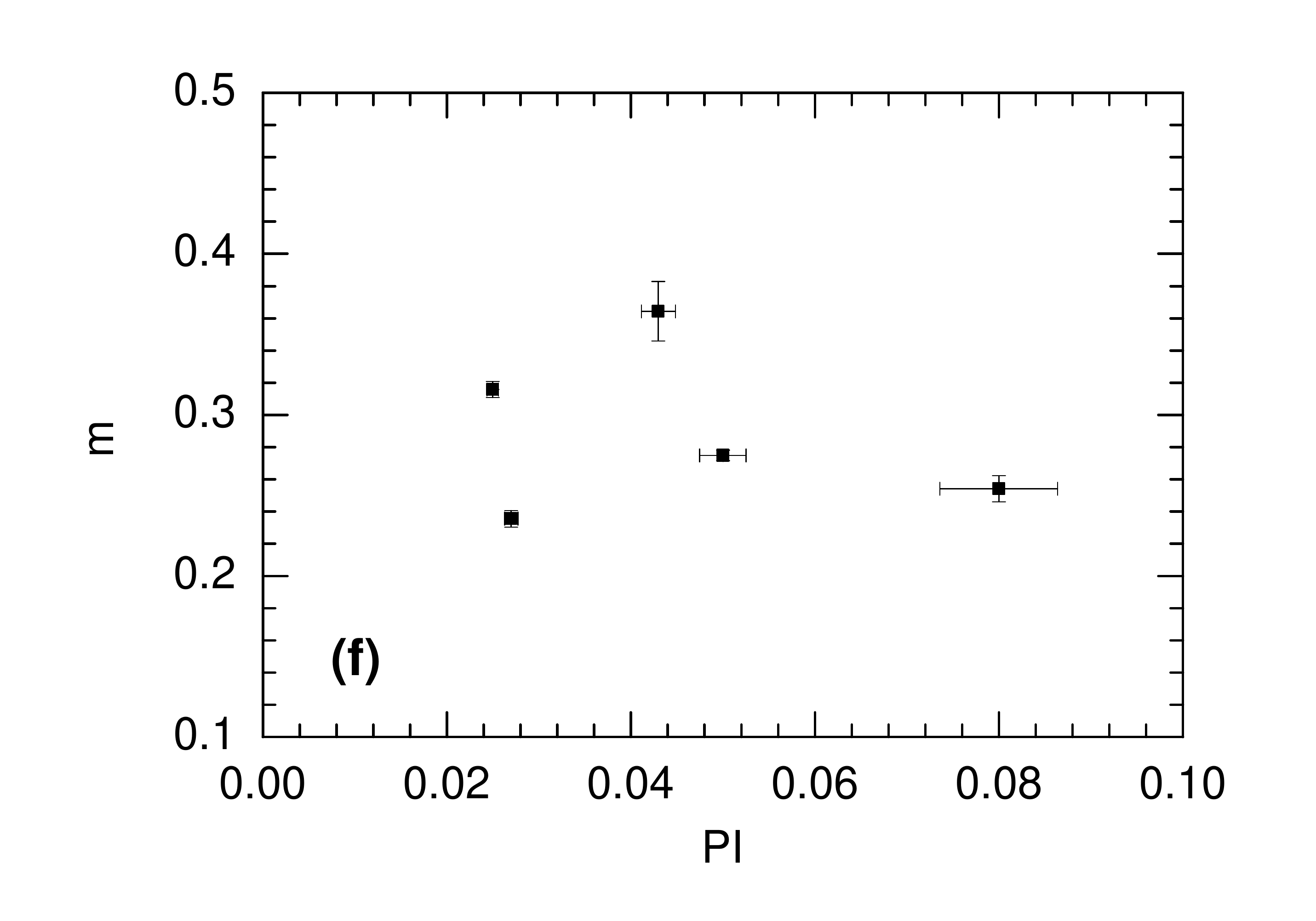}\hspace{0.2cm}}
\caption{\label{fig8} (Color online) Correlations between the fitted slope $m$ and various particle characteristics. (a) effective charges, $_{eff.G}$; (b) the number averaged mean particle diameter, 2a; (c) effective surface potential,  $\Psi_{eff}$; (d) effective temperature, ${T_{eff}=k_BT/V(d_{NN})}$, at melting (e) coupling strength at melting, $\kappa d_{NN}$; and (f) polydispersity index, PI. Only in (e), a weak correlation is observed with a slightly negative slope and a correlation coefficient of $r = 0.53$. Note that in (f), $m$ does not vary with PI.}
\end{center}
\end{figure*}

Fig.~7a shows that the equilibrium reduced IFE is not correlated to the slope $m$. This is expected, since also in Fig. 6 and the Turnbull plots found in the literature, $C_T$ is constant irrespective of the measured or calculated $\sigma_0$ \cite{Turnbull JCP 1949, Turnbull JAP 1950, Kelton Solid state Phys 1993, Laird JCP 2001, Hoyt MRS 2004, Jiang rev Surf Sci Rep 2008}. Fig.~7b shows the correlation between the entropy of freezing and the polydispersity index. Also here we observe a clear decrease with increasing PI. Note, however the larger error bars for the cases where $\Delta \mu$ was obtained using the approximation of Aastuen. Its uncertainty translates to an uncertainty in slope and thus in $T\Delta S_f$.

We also checked for any correlations of $m$ to the strength and range of interaction in Figs.~8 a-f. Again the result is negative, possibly with the exception of the weak trend of a decreasing m with increasing coupling parameter $\kappa d_{NN}$ (correlation coefficient $r = 0.53)$.

\bigskip

\textbf{Appendix B: Characterization of particle interactions under deionized conditions}

Supplied suspensions were first diluted and stored over mixed bed ion exchange resin (Amberlite, Rohm \& Haas, France), for a few weeks under occasional gentle stirring. They were then filtered to remove dust, resin debris and coagulate, regularly occurring upon first contact with the exchange resin. The procedure was repeated using fresh resins. All further conditioning was performed in a closed $Teflon^{(R)}$ tubing system containing a column filled with mixed bed ion exchange resin, a reservoir under inert gas atmosphere, to add particles, water or electrolyte, a cell for static light scattering to control the particle number density, $n$, a cell for \textit{in situ} conductivity measurements to control the electrolyte concentration and the actual measuring cell for the crystallization experiments. This procedure allows for a fast and effective deionization and homogenization of the samples. Furthermore it leaves crystallizing suspensions in shear-molten state, from which they readily nucleate and grow crystals, after the shear is stopped. Also the silica species Si77 was first thoroughly deionized, then filled into the circuit and diluted to the desired concentration. NaOH was added up to the equivalence point to obtain maximum charge \cite{Herlach JPCM 2011 colloids as models, Herlach Rev EPJST 2014}.

Under such low salt conditions, van der Waals attraction can be neglected and the pair interaction relevant during solidification experiments is assumed to be a purely repulsive hard-core Yukawa (HCY) potential \cite{Non-DLVO}. Effective electrokinetic charge numbers, $Z_{eff,\sigma}$, were determined from the linear particle number density dependence of the conductivity interpreted in terms of a Drude type model \cite{Hessinger PRE 2000, Medebach JCP 2005}. These agree well with the charge numbers derived from electrophoresis experiments and charge numbers obtained from the fit of a screened Coulomb potential to the numerical solution of the non-linearized Poisson-Boltzmann equation within a cell-model \cite{Shapran CSA 2005}. The reduction of the effective conductivity charge compared to the bare charge is a consequence of the so-called counter-ion condensation and in fact gives a measure of the number of freely moving counter-ions \cite{Levin RPP 2002}. Furthermore, effective elasticity charges, $Z_{eff,G}$, were derived from shear modulus measurements on polycrystalline samples using torsional resonance spectroscopy and interpreting the obtained data in terms of an effective HCY pair potential \cite{Wette CSA 2003}. In addition to the effects of counter-ion condensation, this effective charge also accounts for many-body terms in the potential of mean force, the so-called macro-ion shielding \cite{Klein JCPM 2002}. The latter effect is not present for isolated pairs but starts as soon as a third particle is present \cite{Russ EPL 2005, Kreer PRE 2006}. We note that both effects are due to the overlap of particle electric double layers and that they tend to be fully developed in the case of counter-ion dominated screening. For highly charged particles, as used here, counter-ion dominated screening already develops for densities well below the freezing density and therefore is the case in all crystallization measurements. It is worth mentioning that this condition also ensures, that slight errors in the deionization control will have only a marginal effect on the pair interaction, because the counter-ions provided by the particles themselves by far outnumber any residual electrolyte ions (c.f. Eqn. B.2).

In general, the two effective charges, $Z_{eff,\sigma}$ and $Z_{eff,G}$, differ by some 40\% \cite{Wette CSA 2003}. Further, the charges are close to the theoretically expected saturation limit $Z_{eff,i} = \Psi_{eff,i} a/\lambda_B$ with $\lambda_ = 0.72nm$ being the Bjerrum length, and $\Psi_{eff,i}$ being the slightly surface chemistry dependent effective surface potential measured by different techniques, $i$ \cite{Gisler JCP 1994}. We found, that utilizing $Z_{eff,G}$ to localize the observed melting line in the effective temperature - coupling parameter plane of the phase diagram regularly yielded a good agreement of our results with the theoretical predicted location of the melting line \cite{RKG JCP 1988, Hamaguchi PRE 1997, Smallenburg JCP 2011, Hynninen PRE 2002, Hynninen PRE 2004, Hynninen PRE 2005, Hynninen JPCM 2003}. By contrast, no agreement was observed when $Z_{eff,\sigma}$ was used \cite{Lorenz JCP 2010, Wette JCP 2010 PDG Si, Wette PCPS 2006 Consistence, Lorenz JPCM 2009}, i.e. when neglecting many-body effects on the effective charge. We therefore use $Z_{eff,G}$, the particle number density $n$ and the micro-ion number density, $n_s$, as input for calculating the hard-core Yukawa pair interaction energy in the present study:

\begin{equation}\tag{B.1}
V(r)=\frac {Z_{eff,G}^2e^2}{4\pi\epsilon} ({\frac {\exp(\kappa a)}{1+\kappa a}})^2 \frac {\exp (\kappa r)}{r}
\end{equation}
 	
with the elementary charge, $e$, the solvent dielectric permittivity  ${\epsilon=\varepsilon_0\varepsilon_r}$, and the screening parameter \begin{equation}\tag{B.2}
\kappa = \frac {e^2}{\epsilon {k_BT}} \sqrt{n Z_{eff,G}z^2 +n_sz^2}
\end{equation}

where $z = 1$ is the micro-ion valency. The micro-ion number density $n_s$ is calculated accounting for ions stemming from added electrolyte (as measured from conductivity) and dissolved CO$_2$ (using temperature dependent solubilities \cite{Millero CO2}) as well as ions from the self dissociation of the solvent ($c = 2$ $10^{-7} mol l^{-1}$ at pH 7). Note that for the analyzed data, the counter-ion density which is explicitly accounted for through the $nZ_{eff}$-term  in all cases contributes the overwhelming majority of screening ions .

\bigskip

\textbf{Appendix C: Determination of CNT-based effective non-equilibrium IFEs from nucleation and growth measurements}
\bigskip

To apply CNT for obtaining estimates of the effective non-equilibrium IFEs one needs nucleation rates measured at known meta-stability. The procedures involved to obtain both rates and the corresponding $\Delta\mu$-values from optical experiments have already been described in the literature (\cite{Palberg JPCM Rev 1999, Palberg JCPM Rev 2014} and references therein). We here outline both. We start with the growth measurements used to determine $\Delta\mu$ and continue with the nucleation experiments to obtain $J$. We the proceed with a comparison of CNT-based evaluation schemes to obtain estimates of $\gamma$ and $\sigma(\Delta\mu)$.

\begin{figure}[htb]
\begin{center}
\includegraphics[width=0.5\textwidth,angle=0]{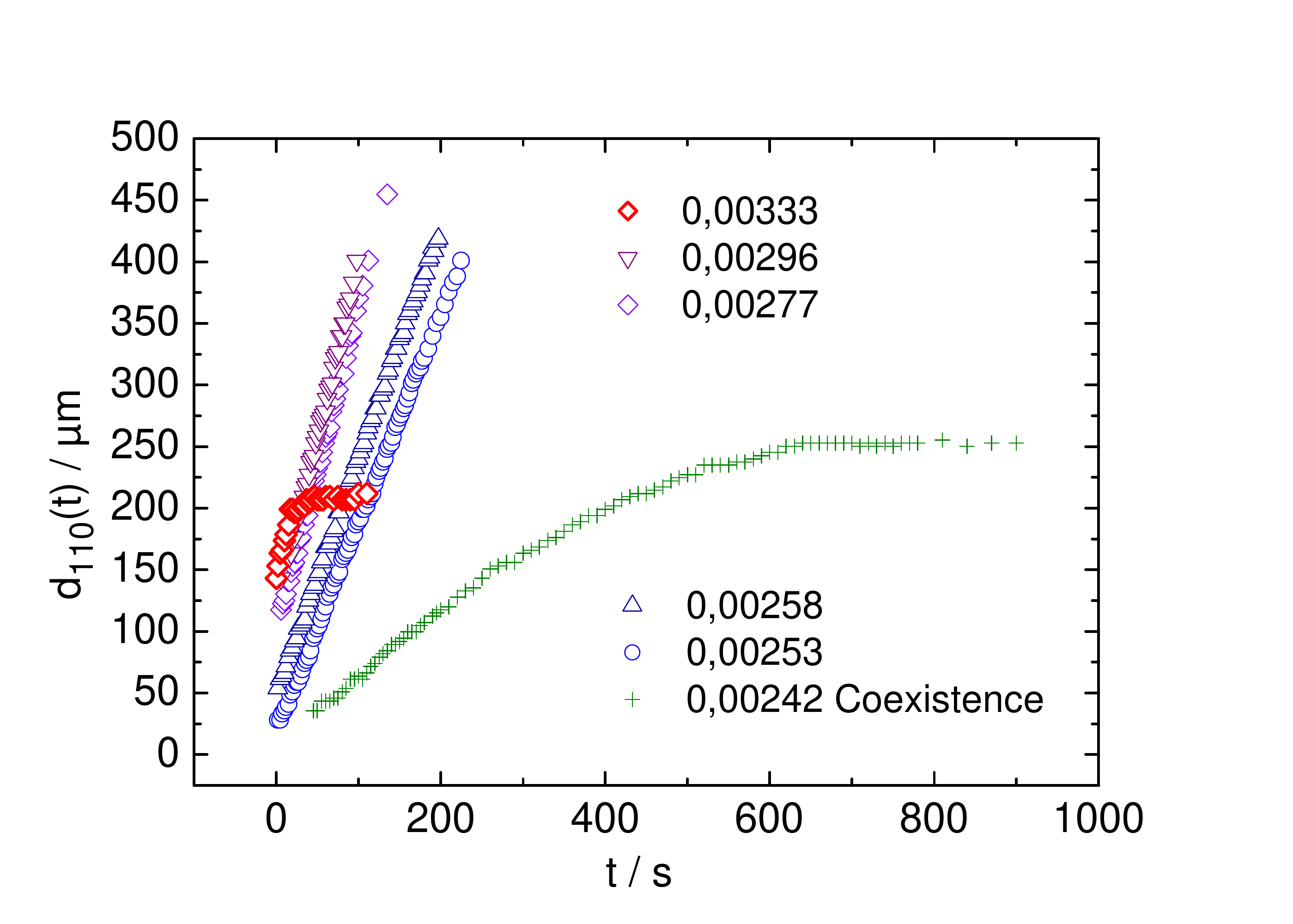}\hspace{0.5cm}
\caption{\label{fig9}(Color online) Typical wall crystal growth curves obtained for different particle volume fractions as indicated.}
\end{center}
\end{figure}

Growth measurements employ microscopy to obtain the growth velocity of crystals. Most suitable contrast variants are Bragg microscopy \cite{Wurth PRE 1995} or polarization microscopy \cite{Monovoukas Langmuir 1991}. Both allow direct determination of the crystal extension from the microscopy images. Rectangular flow through cells are used for microscopic investigation. Charged spheres crystallize with bcc structure at low salt and particle concentrations with a narrow coexistence region separating freezing and melting line in a particle number density \textit{versus} salt concentration phase diagram \cite{Palberg JPCM Rev 1999}. Using the conditioning system described above, the suspension is kept in a shear molten state. The flat cell wall acts as nucleus for a wall crystal which starts growing in [110] direction immediately after cessation of shear \cite{Stipp JPCM 2004}. Just above melting, growth in [110] direction is slower than the average radial growth, but this difference vanishes for larger meta-stability \cite{Wurth PRE 1995}. Experiments are best performed at conditions just above melting. Across the coexistence region, a sub-linear growth is observed, due to the parallel establishment of the difference in density between both phases. Far above melting, growth occurs over very short times only, as it gets very fast and is quickly stopped upon intersection with bulk nucleated crystals. Just above melting, linear wall crystal growth is observed over sufficiently extended times and lengths. Fig. 9 shows typical growth curves.  Growth velocities above coexistence (at coxistence) are inferred from the slope (the limiting slope for $t=0$) of the curves. They first increase with increased meta-stability but then level off at a plateau. Such behaviour is typical for reaction controlled growth and well described by a Wilson-Frenkel growth law:

\begin{equation}\tag{C.1}
v(\Delta\mu)= v_\infty (1-exp(-\Delta\mu / k_BT))
\end{equation}

with the limiting velocity, $v_\infty$. The crucial point is the use of a suitable approximation for $\Delta\mu$, the chemical potential difference between the two phases. In principle, this quantity depends on the independently measurable interaction parameters effective charge, $Z_ {eff}$ (from elasticity measurements \cite{Wette CSA 2003}), particle number density, $n$ (from static light scattering \cite{Luck Natwiss 1963, Palberg JPCM Rev 1999}) and salt concentration, $c$ (from conductivity \cite{Medebach JCP 2005, Hessinger PRE 2000, Wette CSA 2003}). In his seminal work, Aastuen et al. suggested to use the approximation:

\begin{equation}\tag{C.2}
\Delta\mu \simeq B \frac{n-n_F}{n_F}
\end{equation}

where $F$ denotes feezing and B is a proportionality constant used as second fit parameter. This approximation neglects any influence of the charge and the salt concentration and any change of the interaction with $n$. W\"urth et al. \cite{Wurth PRE 1995} therefore suggested using a reduced density difference
\begin{equation}\tag{C.3}
\Delta\mu \simeq B \Pi^* = B \frac{\Pi-\Pi_F}{\Pi_F}
\end{equation}

with ${\Pi=\alpha  n V(d_{NN})}$, $V(d_{NN})$ denoting the pair interaction potential at the nearest neighbour distance and $\alpha$ being a coordination number \cite{Wurth PRE 1995}. Since the latter may differ within different phases, one compares the values for the melt to those of the fluid phase at freezing. This exploits that close to a phase transition the Gibbs free energy difference is approximately linear for any pair of phases. W\"urth's approximations have been thoroughly tested and curves of $v$ versus $\Pi^* $ measured varying different interaction parameters collapse to a single master curve \cite{Palberg JPCM Rev 1999} which is well described by Eqn. (C.1). It thus accounts for changes in any of the interaction parameters and allows, for instance, measurements in dependence on the salt concentration at fixed $n$. Typical fit parameter values are $B = (1.5-15)$ and $v_\infty = (2-20)\mu m/s$. Limiting velocities show some scaling with the particle size \cite{Palberg JPCM Rev 1999}, but there are indications that in addition the thickness of the interfacial region may be of some importance \cite{Stipp PhilMag 2007}.

\begin{figure}[htb]
\begin{center}
\includegraphics[width=0.5\textwidth,angle=0]{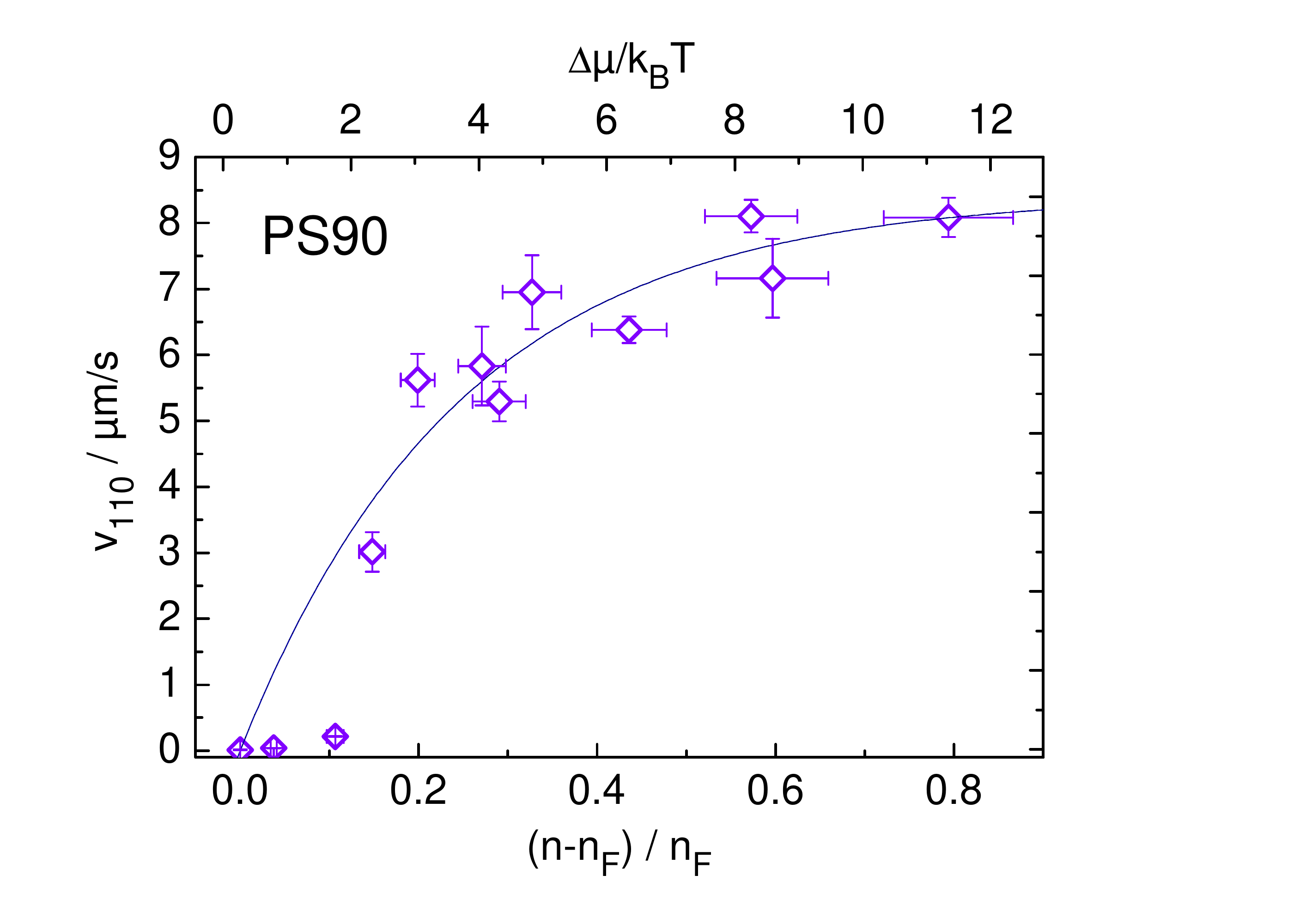}\hspace{0.2cm}
\caption{\label{fig108}(Color online) Growth velocities of PS90 in dependence on the reduced density. The solid line is a fit using Eqn. C.1 with Eqn. C.2. The fit parameters obtained are ${B_{PS90} = (4\pm0.6)}$  for the proportionality constant and $v_\infty = 8.4\pm0.3 \mu m /s$ for the limiting velocity.}
\end{center}
\end{figure}

At the time of the original publication of the nucleation data of PS90, PS100B and their mixture, no growth data was available. Some time later, these were measured and our results for PS90 are shown in Fig. 10. Note the low velocities across the coexistence region ${n_F\leq n \leq n_M}$ with $n_M$ now determined to be ${n_M = 4.6\mu m^{-3}}$. Unfortunately, the salt concentration was not measured accurately enough in these experiments and therefore the fit of the WF-law could only be performed using Aastuen's approximation. It returned a value of ${B_{PS90} = (4\pm0.6)}$  for the proportionality constant and $v_\infty = 8.4\pm0.3 \mu m /s$ for the limiting velocity. The growth data for PS100B was measured by Liu et al. \cite{Liu JCP 2005} with simultaneously determined salt concentrations. Evaluation using Aastuen's (W\"urth's) approximation yielded $B = 4.0\pm0.3$ ($B = 2.6\pm0.2$). For the present paper, we decided to use the value from Aastuen's approximation, which allowed to adopt an estimate of $B = 4.0$ for the mixture as well. The new values for $\Delta\mu$ were used to re-evaluate the nucleation experiments and obtain estimates for the IFE. The corrected values of the IFEs differ only slightly from the original ones, but their incrrease with increasing $\Delta\mu$ was found to be much stronger due to the use of the smaller proportionality constant.

Measurements of the nucleation rates were performed  by microscopy \cite{Wette JCP 2005 Microscopy}, static light scattering or time resolved USAXS measurements \cite{Wette EPL 2003, Herlach JPCM 2011 colloids as models, Wette JCP 2010 PDG Si}. Direct video microscopy could be applied at low meta-stability, where nucleation sites are sufficiently distant to be resolved and nucleation rates are small enough to be followed (typically well below $10^2 s^{-1}$). Rates were divided by a suitable expression for the free volume. The original Avrami model considers bulk nucleating crystals only, assuming their sites to be Poisson distributed and their nucleation rate density being constant in time \cite{Avrami}. It was extended by Wette et al. \cite{Wette JCP 2005 Microscopy} to also include competing wall crystal growth and variable rates. The resulting expression for the relative free volume at time $t$ reads:

\begin{equation}\tag{C.4}
\begin{split}
F(t)& = \frac{(V_0-2Ad_0-2Av_Wt)}{(V_0-2Ad_0)}\times\\
& exp(- \frac{4\pi}{3}\sum_{i}{\frac{m_i}{V_0-2Ad_0-2Av_W\tau_i}[R_0+v{(t-\tau_i)}^3])}
\end{split}
\end{equation}

where $V_0$ is the total observed volume, $A$, $d_0$ and $v_W$ are, respectively, the observed area, initial thickness and growth velocity of the wall crystal. $m_i$ is number of crystallites appearing at times $\tau_i=i\Delta t$ with $\Delta t$ typically on the order of a few tenths of a second. $R_0$ and $v$ are the bulk crystallite radius at first identification and its growth velocity, respectively. An example of resulting nucleation rate densities obtained for PnBAPS68 is semi-logarithmically plotted in the inset of Fig. 11. After a short induction time, $J(t)$ first increased sharply, then settled to a plateau, before decreasing again. With increasing $n$ the plateau extension shrank and the maximum values increased considerably. Close to the phase boundary, the nucleation rate density stayed constant over an extended time and therefore the steady state nucleation rate density, $J_{SS}$, required by CNT, was well approximated by the plateau value $J_{MAX}$. At larger $n$ data was fitted by Kashchiev's theory of transient nucleation \cite{kashchiev}. This is also shown in the inset of Fig. 11. Note the increased value at long times, which is identified to $J_{SS}$ and becomes larger than $J_{MAX}$ for number densities ${n>10^{19}m^{-3}s^{-1}}$.

\begin{figure}[htb]
\begin{center}
\includegraphics[width=0.5\textwidth,angle=0]{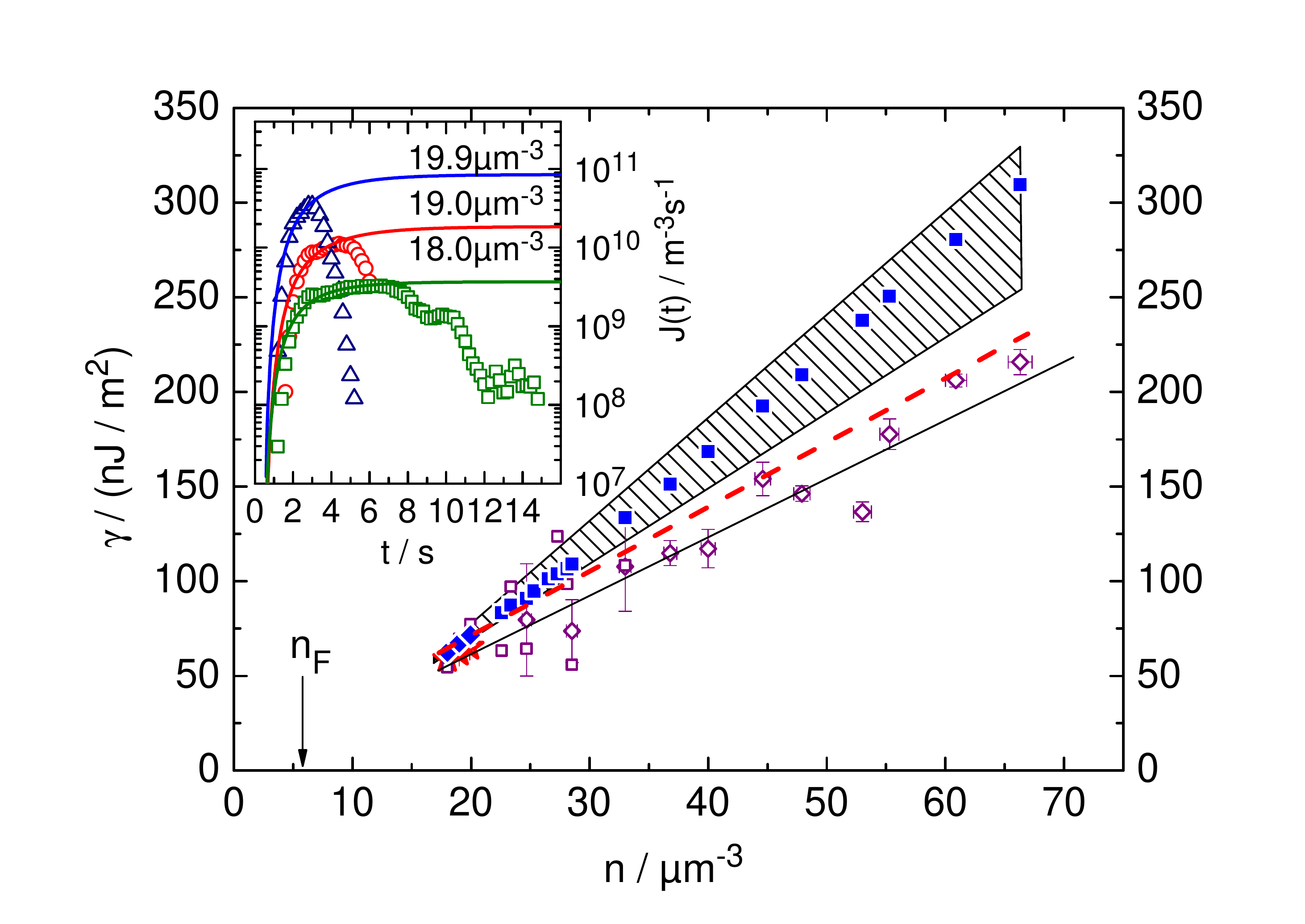}\hspace{0.2cm}
\caption{\label{fig11}(Color online) Density dependence of the interfacial free energies, $\gamma$ of PnBAPS68 obtained from different evaluation schemes. Solid symbols denote use of CNT with $J_0$ calculated using Eqn.~(C.6) in Eqn.~(C.5) with $A=1$ and $D_{eff} = 0.1D_0$ and different input data: average nucleation rate density, $J_{AVR}$  (squares), maximum nucleation rate density, $J_{MAX}$ (diamonds) or steady state nucleation rate density, $J_{SS}$ (stars); hatched area denotes the maximum systematic change by setting $D_{eff} = D_0$ (upper bound) or $D_{eff} = 10^{-3}D_0$ (lower bound). Open symbols denote data from two different experimental runs using a corrected version graphical evaluation \cite{HJS priv Comm}.  The solid line gives a least-square linear fit to these data. The dashed red line gives the $\gamma(n)$ values obtained from the fit of Eqn.~(C8) to the data as shown in Fig.~12. All curves increase linearly with n for $n>n_F$. Inset: time dependent nucleation rate densities for three densities $n$ as indicated. With increased $n$, transient effects become more pronounced. Solid lines are the fits of Kashchiev's theory yielding $J_{SS}$ \cite{kashchiev}.}
\end{center}
\end{figure}

At larger rates, post solidification images of the sample were taken. The distributions of probability density for the radially averaged linear dimension of crystallites, L, are obtained by image analysis. Typical distributions were slightly skewed to large $L$ values and well described by a log-normal distribution. With increasing $n$, the average crystallite size shifted to smaller values. Since here the wall crystal correction to F(t) was negligible, the original Avrami formula was used. It connects the crystallite density and growth velocity to an average nucleation rate density as: ${J_{AVR} = (1/\alpha) v \rho^{4/3}}$. Here, $1/\alpha =1.158$ is a geometrical factor. ${\rho \equiv {\langle L \rangle}^{-3}}$ is the crystallite density and $\langle L \rangle$ is the average linear dimension of crystallites assumed to be cube-shaped.

At still larger rate densities, the crystallites became too small to be properly resolved by microscopy. Then scattering methods were applied. Here, raw data $I(q,t)$ were first de-smeared \cite{Herlach JPCM 2011 colloids as models, Wette EPL 2003} to isolate the scattering signal from apparative contributions. For isolating the crystal structure factor in the time-resolved USAXS measurements, the fluid background (obtained from the first measurement immediately after shear melting) was subtracted with a weighting factor $\beta$ denoting the fraction of remaining melt. We then divided the signal by the independently measured form factor $P(q)$. The isolated crystal structure factor $S_X(q,t)$ showed Bragg peaks which grew and sharpened over time. Further evaluation followed Harland et al. \cite{harland PRE 55} to obtain i) the crystallinity (fraction of crystallized material), $X(t)$, from the integrated intensity normalized to the long time value after complete solidification, ii) the average linear dimension $\langle L \rangle$ of crystallites from the peak width, and iii) their number density, $\rho(t)$, from dividing $X(t)$ by the average volume of crystals. From the derivatives of the average linear dimension and the number density we obtained the growth velocity and the time dependent nucleation rate density $J(t)$. Like for the microscopy data, the time dependent nucleation rate densities could be fitted by Kashchiev's expression for transient nucleation to return $J_{SS}$.

In light scattering experiments, no time dependent data was measured. Rather, a post-solidification analysis was performed, using the average linear dimension $\langle L \rangle$ from the width of the observed Bragg peaks, the limiting velocity from the growth experiments and Avrami's formula to calculate $J_{AVR}$. We note, that the light scattering data may be biased by the presence of a finite and presumably skewed crystallite size distribution. Neglect of this influence in our analysis may, in principle, lead to inconsistencies between the nucleation rates determined by this technique and by the direct size distribution analysis performed at intermediate $n$. In fact, the change of slope in Fig. 12 occurs close to the range in which the data taken by the different techniques overlap. It thus could indicate such an inconsistency. However, this change of slope is systematically present also in Fig. 13. There, the data taken for the other CS samples were taken by light scattering only. Therefore, the good agreement between the data derived from different techniques as displayed in Fig. 12 suggests that a finite crystallite size distribution hardly influences the data evaluation and even less the conclusions drawn in this paper.

PnBAPS68 has been measured using all outlined optical techniques, except USAXS \cite{Wette PhD, Wette EPL 2003, Wette JCP 2005 Microscopy}. In \cite{Wette PRE 2007 CNT} we have compared these to find an excellent agreement between the nucleation rate densities obtained by the different methods. This is also seen in Fig.~12 which shows the nucleation rate densities measured with different methods for $n = (18-67)\mu m^{-3}$. The data covers several orders of magnitude in $J$ without any systematic deviation between the different data sets. Similar data is displayed for a collection of different colloidal species also in Fig. 13. Here, we plotted $J$ \textit{versus} the volume fraction, ${\Phi=\frac{n(4\pi/3)a^3}{V_0}}$ to once more stress the non-space filling character of CS crystals in comparison to HS, which crystallize above $\Phi_M=0.495$ \cite{Zykova JCP 2010}. Note the characteristic shape of all CS curves. The nucleation rate density increases by several orders of magnitude for small changes in $\Phi$. The increase is more pronounced at smaller $\Phi$. By contrast, the HS data displays a maximum, which is attributed to the  vanishing long-time self-diffusion upon approaching the HS glass transition at $\Phi_G = 0.57-0.59$ \cite{Palberg JPCM Rev 1999}.

\begin{figure}[htb]
\begin{center}
\includegraphics[width=0.4\textwidth,angle=0]{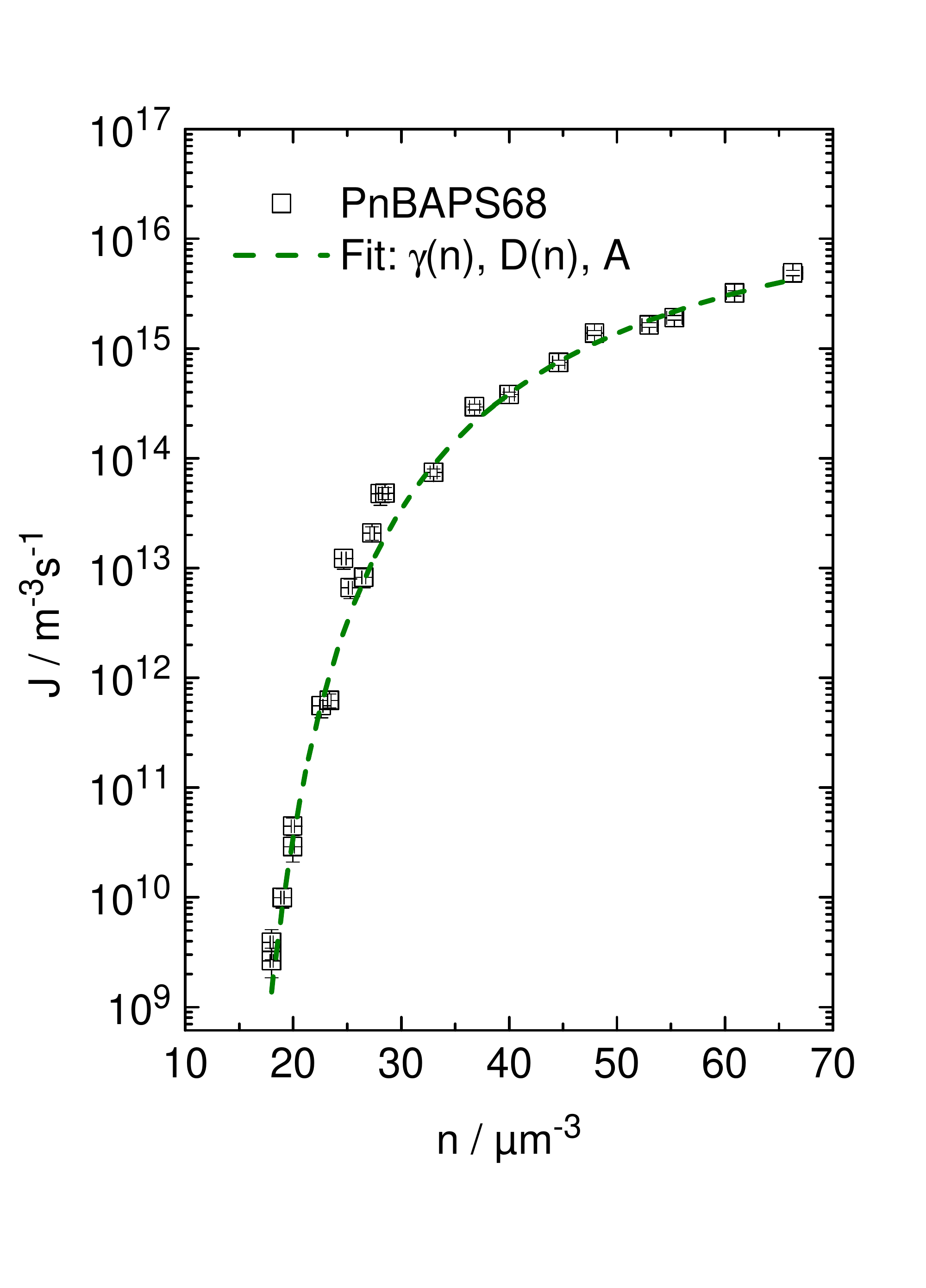}\hspace{0.1cm}
\caption{\label{fig12}(Color online) Nucleation rate densities of PnBAPS68 as measured by video microscopy ($n= (18-19.9)\mu m^{-3}$), post solidification crystal size analysis ($n= (18-33)\mu m^{-3}$) and static light scattering ($n= (25-67)\mu m^{-3}$). Note the excellent agreement between data derived from different methods. The solid line is a least-square fit of Eqn.(C.8) to the data using A, ${D_{L}^S(n)}$ and $\gamma (n)$ as free parameters. An excellent description of the experimental data can be observed. The obtained $\gamma(n)$ are shown in Fig. 11 as dashed red line.}
\end{center}
\end{figure}

While different techniques to obtain nucleation rate densities yield consistent results, systematic differences are introduced in the next step of evaluation. Several CNT-based schemes exist. CNT in its simplest form assumes that the steady state nucleation rate density is given by:

\begin{equation}\tag{C.5}
J_{SS} = J_0 exp(-\Delta G^*/k_BT)
\end{equation}

where for colloidal systems the nucleation barrier ${G^* = 16\pi\gamma^3 /3 {(n\Delta\mu)}^2}$ is determined by the IFE, $\gamma$, the difference in chemical potential, $\Delta\mu$, between the melt and the solid phase and the particle number density, $n$. $J_0$ is a kinetic pre-factor which for colloids with diffusive dynamics and particle by particle attachment was proposed to be \cite{Russel PT (1990), Palberg JPCM Rev 1999}:

\begin{equation}\tag{C.6}
J_0 = An\frac{D_{L}^S}{\ell^2}
\end{equation}

where ${D_{L}^S}$ is the long-time self-diffusion coefficient, A is a dimensionless factor, and $\ell$ a characteristic length scale approximated by ${\ell  = d_{NN}\approx n^{-1/3}}$. In most publications, the dimensionless factor $A$ has been set to unity as a first approximation.

Use of Eqn.~(C.6) in Eqn.~(C.5) to calculate $\gamma$ affords an additional assumption about the $n$-dependence of $D_{L}^S$. For charged spheres showing no glass transition in the range of investigated $n$, $D_{L}^S$ is limited as ${D_0 \geq D_{L}^S \geq 10^{-3}D_0}$, where the upper bound is the Stokes-Einstein diffusion coefficient ${D_0=k_BT / 6\pi\eta a}$ and the lower bound is estimated from the results of published diffusion data on charged spheres \cite{Wagner JCP 2001}. For the calculation of the data points in Fig. 11, we used the approximation $D_{L}^S=0.1D_0$ which corresponds to applying L\"owen's dynamical freezing criterion \cite{Lowen PRL 1995} to estimate the diffusivity at freezing and neglecting the (weak) density dependence over the volume fraction range investigated. The hatched area denotes the maximum systematic change by varying  $D_{eff}$ within the mentioned bounds.

Alternatively, a graphical evaluation from a plot of $ln(J)$ versus ${1/{(n\Delta\mu)}^2}$ was performed \cite{Wette PRE 2007 CNT}. The slope of this curve is ${m = 16\pi{\gamma(n)}^3/k_BT}$. The results of this graphical evaluation are also shown in Fig. 11. They appear to lie systematically below the results of other evaluation schemes. Further, the noise in the $J(n)$ data directly translates into a noticeable scatter of $\gamma(n)$. However, the linear increase with increasing $n$ is clearly seen despite this scatter. Note, that the graphical evaluation does not make any assumptions about the kinetic pre-factor. Thus, the results shown in Fig. 11 demonstrate the pronounced meta-stability dependence of $\gamma$.

Further, an explicit calculation of the kinetic pre-factor within the framework of CNT following \cite{Zeldovic JETP 1942} was performed in \cite{Wette PRE 2007 CNT} to yield:

\begin{equation}\tag{C.7}
J_{0,CNT} = 12 {\left(\frac{4}{3}\right)}^{2/3} \pi^{-1/3} n^{4/3}\sqrt{\frac{\gamma}{k_BT}}D_{L}^S
\end{equation}

In comparison to Eqn.(C.6), this pre-factor has a differing $n$-dependence and further depends on an $n$-dependent non-equilibrium IFE. In \cite{Wette PRE 2007 CNT}, this approach was used to obtain estimates for both the non-equilibrium IFE and the kinetic pre-factor. Interestingly, no acceptable fit could be obtained using Eqn.~(C.7) directly in Eqn.~(C.5). Therefore the authors performed a least-square fit to the $n$-dependent measured nucleation rate densities using

\begin{equation}\tag{C.8}
J_{SS} = An^{4/3}\sqrt{\frac {\gamma}{k_BT}}D_{L}^S exp \left( \frac{-16\pi\gamma^3}{3 k_BT {(n\Delta\mu)}^2}\right)
\end{equation}

Here, a constant $A$, a variable surface tension ${\gamma (n)}$ and a variable self diffusion constant ${D_{L}^S(n)}$ were used as fitting parameters. As shown in Fig. 12, an excellent fit can be obtained. Further, the results of this fit can be described in terms of second order polynomials ${\gamma (n) = (b_0 + b_1n +b_2n^2)}$ and ${D_{L}^S(n) = (a_0 + a_1n +a_2n^2) D_0}$. The polynomial for $\gamma(n)$ is shown in Fig.~11 as dashed red line lying between the results of the other two methods of evaluation.

Taking the reduced values and applying our extrapolation scheme to each, the differently obtained $\gamma(n)$ in Fig. 11 yield estimates for the equilibrium IFE differing by less than 5\%. Slopes and thus Turnbull coefficients differ by approximately 10\%. Even though these differences are small, the results from graphical evaluation appear to be at systematically lower values, while the results from fits of Eqn.(C.5) and (C.6) using $D_{L}^S = 0.1D_0$ systematically show larger values than those obtained \textit{via} fits of Eqn.~(C.8). Therefore, $\sigma(n)$ for all latex spheres compiled in Fig.~1 data were based on $\gamma$ derived \textit{via} fits of Eqn.~(C.8).

\begin{figure}[htb]
\begin{center}
\includegraphics[width=0.5\textwidth,angle=0]{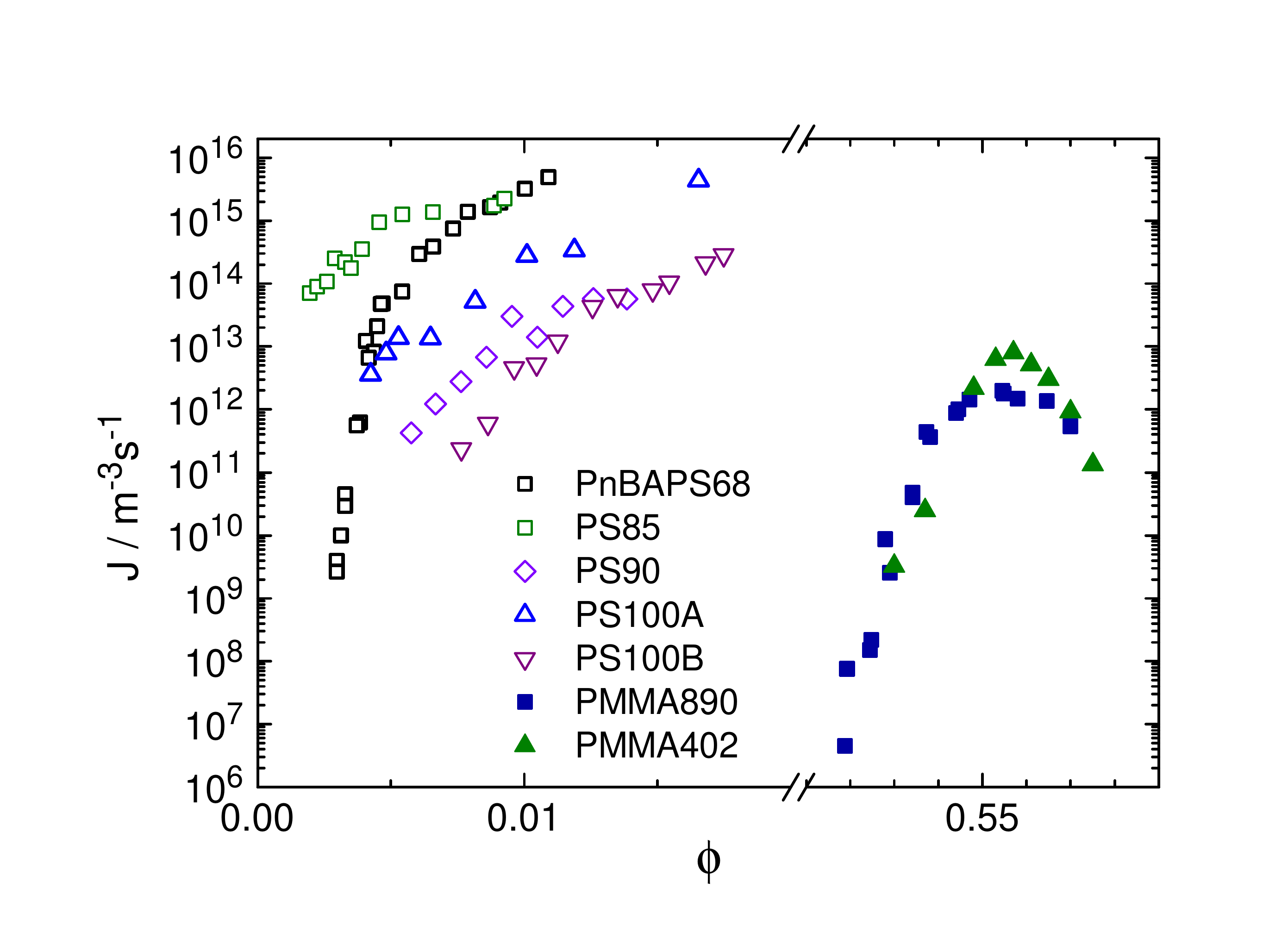}\hspace{0.2cm}
\caption{\label{fig13}(Color online) Nucleation rate densities of different colloidal species versus volume fraction. Open symbols denote charged sphere systems with their diameters indicated. Closed symbols refer to data obtained for two HS systems (PMMA890: \cite{Palberg JPCM Rev 1999}; PMMA402: \cite{harland PRE 55}. Note the low volume fractions of CS and the steepness of the increase in $J$ }
\end{center}
\end{figure}

An equally satisfying agreement is not observed for the kinetic pre-factors obtained from the three different evaluation schemes. This was shown in \cite{Wette PRE 2007 CNT}, too. Using the graphical evaluation, the kinetic pre-factor $J_0, graph(n)$ was obtained from extrapolating the locally fitted slope to ${1/{(n\Delta\mu)}^2}=0$ corresponding to infinite meta-stability. Using Eqn. (C.5) in Eqn. (C.6) with A = 1 results in a prefactor $J_{0,approx}(n)$. Finally, $D_{L}^S(n)$ and $\gamma(n)$ from the fit of Eqn.~(C.8) were used in Eqn.~(C.7) to calculate the $n$-dependent kinetic pre-factor ${J_{0,CNT}(n)}$ in a point-by-point manner. The results for these pre-factors differed considerably. Compared to the measured $J(n)$, $J_{0,graph}(n)$ showed a somewhat less pronounced n-dependence and up to two orders of magnitude larger values. The difference was getting smaller with increasing $n$. The n-dependence of both other sets of pre-factors was by far weaker. I.e. in both curves, an increase of $J_0(n)$ of about one order of magnitude over the complete investigated range of $n$ was observed. However, $J_{0,approx}(n)$ was found to be up to seven orders of magnitude larger than measured $J(n)$ and $J_{0,CNT}(n)$ was found to be up to twelve orders of magnitude larger than $J(n)$. Thus far, no good reason has been proposed, as to why the kinetic pre-factors show such large discrepancies, while the IFEs appear to be rather insensitive to the choice of the evaluation scheme.

\bigskip

\textbf{Appendix D: Use of CNT and related schemes to obtain non-equilibrium IFEs}
\bigskip

CNT provides a simple model for nucleation kinetics in first order transitions \cite{Volmer ZPhys 1926, Kaischew ZPC 1934, Becker An Phys 1935, Zeldovic JETP 1942, Turnbull JCP 1949}. The model is based on the idea that nucleation is an activated process and hence contains a Boltzmann factor with an energy barrier $\Delta G^*$. Further, a kinetic pre-factor limits the reaction rate. Expressions for this pre-factor have been worked out assuming particle-by-particle attachment. CNT was originally proposed to described vapour condensation. Therefore, in most versions of CNT (including those used for melt crystallization), it is further assumed, that the nuclei are spherical. This, in turn, implies that the barrier can be written in terms of surface energy loss and volume energy gain based on a single geometrical parameter only, the nucleus radius $r$. CNT is generally believed to capture the basic physics of homogeneous nucleation. However, the justification of many of the made assumptions is still discussed controversially. Therefore, the desire for a comprehensive microscopic theory remains urgent.

CNT allows simple predictions that only require a few (measurable) bulk thermodynamic data. Its predictive success, however, has been shown to be quite limited. Deviations in estimated and measured nucleation rate densities of up to 35 orders of magnitude occur for melt crystallization of metals as well as for vapour condensation and also in colloidal HS systems \cite{kelton and grier book, Katz PAC 1992, Fladerer JCP 2006, Auer Nature 2001 HS mono, Palberg JCPM Rev 2014}. Also other measured quantities deviate strongly from predictions \cite{Feldmar AIP 2013}. This has been blamed on a number of issues, both conceptually and practically. Finding workarounds or solutions has become a field of great interest \cite{kelton and grier book, Das Book}.

The most obvious conceptual criticism aims at the use of a macroscopic concept, the IFE, on the scale of clusters and the application of equilibrium values of thermodynamic quantities under non-equilibrium conditios. In fact, CNT takes a macroscopic continuum view and describes clusters of discrete particles as small, non-interacting chunks of thermodynamically well defined new bulk phases separated from unchanged, isotropic and homogeneous background melt. Discreteness enters only through growth by addition of individual particles. This view implies a number of consequences. For example, CNT assumes nuclei with sharp interfaces. As already early simulation work has revealed \cite{Oxtoby JCP 1982, Haymet JCP 1982}, this capillarity approximation is hardly ever met on the molecular scale. Like for equilibrated, flat interfaces, Density Functional Theory and explicitly microscopic models of the interface have therefore been used to meet this challenge \cite{Das Book, Spaepen Acta Metallica 1975}. In addition, in CNT, nuclei are often assumed to be spherical. Typically, this is an over-simplification, as both computer studies and experiments on metals or colloids show \cite{Schilling PRL 2011, Tan Nat Phys 2014, bokeloh PRL 2015}. Based on Walton's  Atomic Nucleation Theory (ANT) \cite{Walton JCP 1962} and other approaches, several variants have been developed to include non-spherical shapes into the CNT scheme or presenting corrected CNT formulas \cite{Kashchiev JCP 2008}. Alternatively, the spherical shape was kept and a curvature and thus size dependent IFE was introduced \cite{Troster JCP 2012, Wilhelmsen JCP 2015}. Both attempts to describe the very small nuclei encountered at large meta-stability. Under conditions of large meta-stability, another issue is encountered, because due to its construction CNT and extensions do not allow for a description of spinodal processes \cite{Wu Solid State Phys. 1995}. Here, a workaround may be seen in the kinetic model of Dixit and Zukoski \cite{dixit PRE 2001} which in fact yields a good description of the HS nucleation rates down to volume fractions close to freezing.

Furthermore, there are several practical issues. Again, some are concerned with the nucleus shape and size. Scattering experiments typically determine orientationally averaged data referring to an effective sphere of equivalent size. Microscopy reveals irregular and non-spherical forms but often lacks the statistics needed to obtain thermodynamically meaningful averages. Further, the assumption of spatial and temporal homogeneity in terms of pressure, temperature and concentration of monomers may not be met, i.e. in experiments very often it cannot be assured that conditions stay that way throughout the complete crystallization process \cite{Herlach IMR 1993}. This affords additional theoretical efforts \cite{kashchiev, Reguera JCP 2003 I, Reguera JCP 2003 II}. In addition, at large meta-stability, interactions between neighbouring nuclei may occur leading to jamming or coalescence \cite{Schope JNCM 2002} and two-step nucleation may be an important alternative mechanism \cite{Franke SM 2014, Vekilov CGD 2004, Gebauer CSRev. 2014}. Finally, issues of constrained volume may play a role in both experiment and simulations \cite{Schrader PRE 2009}.

In view of all these issues, the mediocre performance of CNT and its derivatives in \textit{predicting} nucleation rate densities, critical nucleus sizes or onset of nucleation with decreasing temperature or increasing pressure is not very surprising. As discussed above, this is also observed in the data on PnBAPS68. Using Eqn.~(C.7) in Eqn.~(C.5) no acceptable two parameter fit to the nucleation rate densities of PnBAPS68 could be obtained. Further, the pre-factors calculated using Eqn.~(C.7) with the results of the very good fit of Eqn. (8) to the data shown in Fig.~12 appeared to be unreliably large values.

However, in the present paper and the ones providing the original IFE data, CNT was used differently. In fact, it was only utilized to parameterize nucleation kinetic data and to obtain estimates of the non-equilibrium IFE as a function of system meta-stability. This use of CNT is much less ambitious and is by far less challenged by the aforementioned issues. For instance, the extended experiments on PnBAPS68 using several different techniques ranging from direct counting to post-solidification analysis employing Avrami theory yielded consistent nucleation rate densities with $J(n)$ overlapping over more than an order of magnitude in $J$. As seen in Fig.~13, this data is representative for a large number of CS systems and covers nucleation rate densities over many orders of magnitude. Three different CNT-based methods of extracting IFEs were employed. The graphical evaluation results are of particular importance. As can be seen in Fig.~11 the effective IFEs assemble within statistical uncertainty on a straight line which demonstrates the linear dependence of the non-equilibrium IFE on the degree of meta-stability. Since this result was obtained without making any assumption on the kinetic pre-factor it lays the basis for the validity of the extrapolation procedure employed in the main part of this paper. The same trend is seen for the data derived using Eqn.~(C.6) which neglecs the $\gamma$-dependence of the pre-factor altogether and the IFEs derived from the fit by Eqn.~(C.8), where it was accounted for without explicitly calculating $A$. The comparison in Fig.~11 therefore shows that all three procedures capture the IFE-dependence of the barrier qualitatively correct. The spread in extrapolated equilibrium IFEs due to the use of different evaluation procedures amounts to some five percent which is on the same order than the statistical uncertainty. The spread in slopes is somewhwat larger but still acceptable.

Two more points deserve further attention. First, the low absolute values of the IFE of a few hundred $nJ/m^{-2}$ as compared to e.g. metals, where typically values of about $250mJ/m^{-2}$ are obtained \cite{Jiang rev Surf Sci Rep 2008, bokeloh PRL 2015}. This is due to the low number density of colloidal suspensions and has the important consequence that colloidal crystal nuclei have fuzzy shapes \cite{Tan Nat Phys 2014, Kratzer SM 2015}, while metal crystal nuclei appear to be much more compact \cite{bokeloh PRL 2015}. The reduced values for the CS IFE compared in Fig.~6, however, are only an order of magnitude lower than the reduced metal IFEs. Second, The different treatment of the CNT-predicted $\gamma^{1/2}$-dependence of $J_0$ in the three evaluation schemes results in systematic quantitative differences in the derived $\gamma(n)$. We here opted for using a fit procedure for all latex systems, because it incorporates the full CNT-predicted IFE-dependence of $J$, had the least statistical and/or systematic uncertainties and shows values midway between those resulting from the alternative schemes. The obtained values are therefore understood as CNT based effective non-equilibrium estimates of the IFE.

In the main part of the paper we extrapolated this data to zero meta-stability. This effort is not backed by CNT itself. Rather it is suggested by the linear dependence of $\gamma$ on $n$, respectively the linear dependence of $\sigma$ on $\Delta\mu$ up to largest meta-stabilities as seen in Fig.~1. We note that for PnBAPS68 the radius of the critical nucleus varied over the investigated range between one and several $d_{NN}$ \cite{Wette PRE 2007 CNT}. The other species had been investigated at even larger meta-stability. The observed strictly linear dependence of $\sigma$ on $\Delta\mu$ thus appears to exclude any dependence of $\sigma$ on the size of the critical nucleus. A linear dependence of the non-equilibrium IFE on meta-stability has been observed before in many systems and was extensively discussed e.g. by Jiang \cite{Jiang rev Surf Sci Rep 2008}. Moreover, in HS, the volume fraction dependent CNT-based estimates of the non-equilibrium IFE linearly decrease with decreasing volume fraction to meet the theoretical and experimental values of the equilibrium IFE at the freezing volume fraction \cite{Franke SM 2014}. Our present extrapolation returned effective equilibrium IFE values between those of HS and those of metals. It remains to be seen, whether future simulations or measurements of the macroscopic IFE of colloidal CS with explicit micro-ions will coincide.

\end{document}